\documentstyle[11pt]{article}     
\begin{document}      
\begin{center}
\large{
The Possibility of Curved Spacetime, Black Holes, and Big Bang is Less than One Billionth }
\normalsize\\
Jin He \\
Department of Physics, Huazhong University of Science and Technology, \\Wuhan, Hubei 430074, China\\
E-mail:mathnob@yahoo.com\\
\mbox{   }      
\end{center}
\large {{\bf Abstract} } \normalsize  
Gravity whose nature is fundamental to the understanding of solar system, galaxies and the structure and evolution of the Universe, is theorized by the assumption of curved spacetime, according to Einstein$^,$s general theory of relativity (EGR). Particles which experience gravity only, move on curved spacetime along straight lines (geodesics). The geodesics are determined by curved-spacetime metric. In the last year, I proposed the mirrored version of EGR, the flat-spacetime general relativity (FGR), in which particles move along curved lines on flat spacetime. This puts gravitational study back to the traditional Lagrangian formulation. The Lagragian on flat spacetime is simply taken to be the curved spacetime metric of EGR. In fact, all acclaimed accurate verification of general relativity is the verification of FGR,  because relativists when confronting GR to observational data, calculate time, distance, or angle by directly using the coordinates in Schwarzschild solution or in post Newtonian formulation. For example, two famous tests of general relativity are about angles. All mainstream textbooks and papers calculate the angles by directly using the coordinate $ \phi $. 
However, only when spacetime is flat does there exist one coordinate system which has direct meaning of time, distance, angle, and {\it vice verse}.
This is the famous Riemann theorem when he pioneered the concept of
curved space. According to the theorem, all coordinates on a curved space are merely parameters. Real angles and distances have to be calculated by employing the coefficients of spatial metric. 
If we do follow the geometry of curved spacetime (EGR) then the deflection of light at the limb of the sun is 1.65 arcseconds (Crothers, 2005). The publicly cited value (1.75 arcseconds) which best fits observational data is predicted by FGR. Therefore, the more claims are made that classical tests of general relativity fit data with great accuracy, the more falsified is the curved-spacetime assumption. That is, the claim is specious to EGR. Relativists made three specious claims as collected in the present paper. 
However, FGR predicts observationally verified results for solar system, galaxies, and the universe on the whole. Because FGR uses the single consistent Lagrangian principle, it is straightforward to show that the possibility of curved spacetime, black holes, and big bang is less than one in billion. An experiment is proposed whose results will completely decide the fate of curved spacetime assumption. Dare to make public the recent result of new Brillet and Hall experiment with one vertical light beam? Note that the original article ``Einstein`s Geometrization vs. Holonomic Cancellation of Gravity via Spatial Coordinate-rescale and Nonholonomic Cancellation via Spacetime Boost`` is attached. \\
\\
Special Relativity --- General Relativity -–- Classical Tests  
\\
\\
{\bf Flat-spacetime Lagrangian vs. Curved spacetime  }\\
\\ 
People found four physical interactions (i.\,e., forces): gravitational, electromagnetic, weak, and strong ones.  
Gravity, the weakest one, is fundamental to the understanding of solar system, galaxies and the structure and evolution of the Universe.
By now, people have suggested two fundamental principles which are used to construct physical theories. One is flat-spacetime Lagrangian principle and the other is  curved-spacetime assumption. Mainstream gravitational theory falls into the latter category (Einstein$^,$s general relativity, EGR) while all other interactions are unified by the former principle.  
A flat spacetime means a global inertial frame. Therefore, Einstein denied the existence of global inertial frames and denied the possibility that our universe might provide the unique inertial frame. 
If Einstein is wrong then all theories of black holes and big bang are wrong, and the theory of gravity simply returns to normal Lagrangian formulation and all four interactions are hopeful to be unified. 

Any theory must be testified. The curved-spacetime assumption, however, is not verified. All hailed accurate verification of general relativity is in fact the verification of flat spacetime,  because relativists when confronting GR to observational data, calculate time, distance, or angle by directly using the coordinates in Schwarzschild metric or in post Newtonian formulation. For example, two famous tests of general relativity are about angles. All mainstream textbooks and papers calculate the angles by directly using the coordinate $ \phi $. However, Riemann proved the famous theorem:
only when spacetime is flat does there exist one coordinate system which has
direct meaning of time and distance, and {\it vice verse}.
That is, all coordinates on a curved spacetime are merely parameters. Real angles and distances have to be calculated by employing the coefficients of  spatial metric. If we do follow the geometry of curved spacetime (EGR) then the deflection of light at the limb of the sun is 1.65 arcseconds (Crothers, 2005). The publicly cited value (1.75 arcseconds) which best fits observational data is based on direct coordinate calculation. 
Therefore, the more claims are made that direct coordinate calculation fits data very well, the more falsified is the curved-spacetime assumption.
That is, the claim is specious to EGR. Relativists made three specious claims as collected in the present paper. 

Now, I review the flat-spacetime interpretation of general relativity (FGR) which I proposed in the last year. Firstly, I introduce the concept of Newton$^,$s as well as Einstein$^,$s inertial frames. 
In Section 2, Einsten$^,$s equivalence principle is demonstrated to be false, the principle being the second specious claim made by relativists. The principle is the only excuse for Einstein to suggest curved spacetime. {\bf In the final part of the Section, I show that the possibility is less than one in billion that the assumptions of curved spacetime, black holes, and big bang are true.} Section 3 presents the third specious claim made by relativists and proposes an experiment whose result will completely decide the fate of curved-spacetime assumption. The final Section is conclusion.

{\bf (i) Newton$^,$s Inertial Frames and his First Law of Motion. }
An inertial frame (flat spacetime) is the one in which the particles which do not suffer any net force are either static or moving in straight lines at constant speeds relative to the frame. This is also called Newton$^,$s first law of motion which can be formulated by an optimization principle (Lagrangian principle). 
The required Lagrangian per unit mass for the particles is the following
$$ 
L\left(\frac{dx ^i}{dt}\right) 
=\frac{1}{2} \left(\left(\frac{dx }{dt}\right )^2 + \left(\frac{dy }{dt}\right )^2 + \left(\frac{dz }{dt}\right )^2 \right)
$$
$$(1)$$ 
where $(x^1=x, x^2=y, x^3=z)$ is the Cartesian rectangular coordinate in the inertial frame. The Lagrangian is the kinetic energy per unit mass of particles.

Now we discuss in inertial frame the motion of particles which do suffer gravity. Firstly, we pay attention to the gravity due to single large point mass $M$ which sits at coordinate origin and is static relative to the inertial frame. Particles no longer move in straight lines at constant speeds and the simple Lagrangian (1) can not describe the motion under gravity. According to Newton theory, the required Lagrangian is the following,
$$ 
L\left(x^i, \frac{dx ^i}{dt}\right) 
=\frac{1}{2} \left( \left(\frac{dx }{dt}\right )^2 + \left(\frac{dy }{dt}\right )^2 + \left(\frac{dz }{dt}\right )^2 \right)   +   \frac{GM}{r}
$$
$$(2)$$ 
where $G$ is the gravitational constant and $r^2=x^2+y^2+z^2$.
Note that spacetime is flat and particles$^,$ motion are curved lines on the flat Euclidean spacetime.
Newton theory explains the solar system very well with very little error.

{\bf (ii) The Difficulty in Newton$^,$s Concept of Inertial Frames. }
The coordinates $t,x,y,z$ are mathematical numbers. How are these numbers achieved? Newton did not give much discussion.   
He required that the number of time were the same for all inertial frames, the universal (absolute) time. In later nineteen century, this assumption led to difficulty in the explanation of light speed. Let us consider two inertial frames, one moving at constant speed $v_0$ with respect to the other. The universal time assumption leads to the conclusion that a light beam at a speed $c_0$ observed by the first frame will have a speed $c_0\pm v_0$ observed by the second one if both the light beam and the frame move in parallel directions.
For example, earth$^,$s orbital speed around the sun is $v_0 \approx $ 30\,km/s. 
Therefore, if we assume that the speed of light observed by solar frame is $c_0$ in the same orbital direction, then the light speed observed by earth frame is $c_0 \pm v_0$.
The difference of the two speeds is   
$$ 
\Delta c = v_0 \approx 30\,{\rm km/s}. 
$$
$$(3)$$ 
Such large difference of light speeds is never observed. Therefore, Newton$^,$s concept of inertial frames must be corrected. Einstein assumes that light speed is the same for all inertial frames.

{\bf (iii) Einstein$^,$s Special Relativity (SR) --- The Innovative Concept of Inertial Frames. }
Therefore, the assumption of single universal time must be abandoned. Time is given by clocks which themselves are physical processes and the physical processes (the clocks) at all places of the frame are static relative to the frame itself. Similarly the rulers which are used to measure spatial distances are static with respect to the frame too. Therefore, the clocks belonging to one inertial frame have relative motion with respect to the clocks belonging to the other one. Therefore, one will find out that the timing of one$^,$s clocks is different from the timing observed by oneself of the clocks in other frame.  Timing and spatial length of a physical process are not universal. If we talk about a time, we need to say by which inertial frame it is given. Therefore, according to Einstein$^,$s special relativity, we have different universal time given by different inertial-frames, instead of single universal time. 

Einstein initiated the special theory of relativity (SR), the new concept of inertial frames which assumed universal value of light speed, instead of universal time. That is, light speed is the same for all inertial frames. Its universal value is
$$ 
c_0= \frac{dx}{dt}= 299,792,458\; {\rm m/s}. 
$$
$$(4)$$ 
This is a theoretical value. Modern technique can measure light speed to the accuracy of decimeters per second directly. However, modern technique can measure the difference of light-speeds of two light beams to the accuracy of $10^{-6}$ meters per second. 
The formula (4) indicates that
$$ 
-c^2_0 dt^2+dx^2=0 
$$
$$(5)$$ 
which suggests a different ``Pythagoras theorem$^{,,}$
$$ 
ds^2=-c_0^2 dt^2+dx^2+dy^2+dz^2. 
$$
$$(6)$$ 
This is called Minkowski indefinite metric on flat Minkowski spacetime, which is the basis of SR. 
Now we need to determine the Lagrangian which describes the motion of particles which do not suffer any force (interaction) in Einstein$^,$s inertial frame.  

The Lagrangian and the light speed are both invariant quantities. Therefore, they must be connected. This is suggested by the above formula (6),
$$ 
\begin{array}{l}
L\left(x^0, x^i, c_0\frac{dt}{dp}, \frac{dx ^i}{dp}\right) \\
=\frac{1}{2} \left( -c_0^2\left(\frac{dt}{dp}\right )^2+ \left(\frac{dx }{dp}\right )^2 + \left(\frac{dy }{dp}\right )^2 + \left(\frac{dz }{dp}\right )^2  \right)
\end{array}
$$
$$(7)$$ 
where $x^0=c_0t$, and $p$ is the curve parameter of particle$^,$s motion. The Lagrangian is called the Lagrangian per unit mass because it does not involve the quantity of mass. 

Because light sets an upper limit for all particles$^,$ speeds, the values of the Lagrangian are not arbitrary. Because we always deal with causal motion, we have $ds^2 \le 0 $, i.\,e.,
$$ 
L\le 0.
$$
$$(8)$$  

Now we derive the Hamiltonian per unit mass for the Lagrangian (7).
The momentums per unit mass canonical to $x^\alpha , \alpha =0,1,2,3$ are the following,
$$ 
\begin{array}{l}
P_0 =\frac { \partial }{\partial (dx^0/dp)} L=-\frac { c_0d t }{dp }  \\
P_i =\frac { \partial }{\partial (dx^i/dp)} L=\frac { dx^i }{dp }, 
\mbox{ }i=1,2,3.
\end{array}
$$
$$(9)$$  
Because the Lagrangian does not depend on time and position coordinates, the momentums are constants, which indicates that the motion of particles governed by the Lagrangian is the one in straight lines at constant speeds,
$$ 
t=a_0p, \; x=a_1p, \;y=a_2p, \;z=a_3p 
$$
$$(10)$$  
where $a_\alpha , \alpha =0,1,2,3$ are constants. 

The Hamiltonian of test particles is
$$ 
\begin{array}{ll}
H &= \frac { dx^0}{dp} P_0 +\sum^3_{i=1}\frac { dx^i}{dp} P_i –- L   \\
&= -\frac{1}{2} c_0^2\left (\frac {dt  }{dp}\right )^2 +\frac{1}{2} \sum ^3_{i=1}\left (\frac {dx^i  }{dp}\right )^2\\
&\equiv L.  
\end{array}
$$
$$(11)$$  
If we choose $a_0 =1$ in (10) then the Hamiltonian (total energy) is
 $$ 
H =E=L=-\frac{1}{2} c_0^2+ \frac{1}{2} \sum ^3_{i=1}\left (\frac {dx^i  }{dt}\right )^2.
$$
$$(12)$$  
Because $E=0$ corresponds to photon$^,$s motion, the formula indicates that light speed is $c_0$ as we expect. 
We see that the spatial part of the Hamiltonian corresponds to kinetic energy while the temporal part corresponds to potential energy.
Both energies are constants. The potential energy is $-c_0^2/2$ which is chosen to be zero in Newtonian theory.   

Einstein$^,$s SR further requires that coordinate transformations between inertial frames are the Lorentz ones and the formulation of all physical laws must be covariant with respect to the transformations. 
Einstein$^,$s SR is verified by many experiments and is the basis of my FGR.

{\bf (iv) Under which Condition is Einstein$^,$s Special Relativity (SR) True? }
Einstein$^,$s special relativity is not true in real condition of local universe. SR is actually the concept of global inertial frames and describes the property of spatially and temporally homogeneous world. It is very important to know that SR would be perfectly and globally true if the matter content of the universe were both spatially and temporally homogeneous. However, it is a fact that the universe is evolving (temporally in-homogeneous). Fortunately, large-scale spatial homogeneity of the universe is observationally proved. Therefore, SR is globally true for any short period of time of the large-scale universe.  

The local in-homogeneous distribution of matter of the universe introduces local spatial in-homogeneity, which is the subject of Newtonian gravity and Einstein$^,$s general relativity. In this case, special relativity must break down. Especially, light speed is anisotropic (not constant). Because gravity is the weakest interaction, the anisotropy of light speed is hard to detect.  

{\bf (v) Einstein$^,$s General Theory of Relativity (EGR). }
Einstein$^,$s SR (7) (or (6)) is the innovated version of Newton$^,$s concept of inertial frames (1). The Lagrangian (2) generalizes (1) to deal with particles$^,$ motion under gravity. Einstein$^,$s general relativity (EGR) which deals with gravity too, does not generalize his SR. SR describes the full property of homogeneous spacetime while gravity introduces inhomogeneity  on spacetime. Therefore, SR must more or less break down in any theory of gravity. Einstein chose to stake at the assumption that SR is perfectly true in any infinitesimal area of spacetime. Accordingly SR can not be perfectly true in any finite area of spacetime. Otherwise, the corresponding Lagrangian would be (7) and no gravity would be present in the area. The unavoidable consequence of Einstein$^,$s choice is that spacetime must curve.       
Therefore, Einstein gave up the global flat Minkowski spacetime by introducing the curved spacetime whose curvature is gravity, and he abandoned Lorentz coordinate transformations by considering all curvilinear coordinate transformations on the curved spacetime. 

Einstein$^,$s assumption of curved-spacetime brings more complexity than truth.
Firstly, curved spacetime is embodied by non-trivial topology. Because topology is a very complicated mathematical subject, most relativists never take a look at it.
Secondly, the concept of curved spacetime is nothing but temporal and spatial in-homogeneity. Therefore, all coordinates on a curved space are merely parameters. Real time and distance have to be calculated by employing  coefficients of the spacetime metric. The calculation of time and distance by employing metric is very complicated too. 
Therefore, all relativists when confronting GR to observational data, calculate time, distance, or angle by directly using the coordinates in Schwarzschild solution or in post Newtonian formulation. 
However, there is the famous Riemann theorem: 
only when spacetime is flat does there exist one coordinate system which has
direct meaning of time and distance, and {\it vice verse}.
Therefore, the hailed accurate tests of GR verified the flat-spacetime interpretation of GR (my FGR). 
The more claims are made that classical tests of general relativity fit data with great accuracy, the more falsified is the curved-spacetime assumption. That is, the claim is specious to EGR.

{\bf (vi) Flat-spacetime Interpretation of Schwarzschild Metric. }
In the last year, I proposed the mirrored version of EGR, the flat-spacetime general relativity (FGR) in which particles move along curved lines on flat spacetime. This puts gravitational study back to traditional Lagrangian formulation. 
For example, Schwarzschild metric of single point mass is,  
$$ 
\frac {1}{2}\left( \frac{d s}{dp}\right )^2 = L\left(x^0, x^i, \frac{dx ^0}{dp} , \frac{dx ^i}{dp}\right) = -\frac{1}{2} B(r ) \left(\frac{c_0dt}{dp}\right)^2 
 + \frac{1}{2}A(r ) \left(\frac{dr}{dp}\right)^2  +\frac{1}{2}r^2 \left(\frac{d\phi}{dp}\right)^2 ,
$$
$$(13)$$  
where
$$ 
B(r )=1-\frac{2r_g}{r}, \;\; A(r )=\frac{1}{B(r )}=\frac{ 1}{1-2r_g/r } 
$$
$$(14)$$  
and the constant
$$ 
r_g=GM/c_0^2 
$$
$$(15)$$  
is the Schwarzschild radius.
The Schwarzschild metric (13) on curved spacetime is simply taken to be the Lagragian on flat spacetime. 
Although the background spacetime is flat and Cartesian coordinates have direct meaning of time and distance, one of the fundamental assumptions of SR breaks down globally. That is, light speed varies with spatial position and spatial direction as indicated in the following.
However, it is still the maximum one at each position and in each direction.
The Lagrangian (13) is the generalization to the one of no-interaction (7).
I call this type of Lagrangian by homogeneous Lagrangian because it is a homogeneous order-two form of the components of the generalized particle velocity. The value of the Lagrangian is negative so that it describes causal motions of material particles. It can be zero and describes the motion of light.

According to the optimization principle, test particle$^,$s motion follows the corresponding Lagrange$^,$s equations. 
The solution of the Lagrange$^,$s equations is
$$ 
\frac {dt }{dp}=\frac{1}{ B( r)}, 
$$
$$(16)$$  
$$
r^2\frac {d\phi }{dp}=J \:({\rm constant}),
$$
$$(17)$$  
$$ 
\frac{1}{2}\left (\frac{A( r)}{B^2(r )} \left (\frac {dr }{d t}\right )^2+\frac {J^2 }{ r^2}-\frac {c_0^2}{B( r)}\right )=\tilde E \:({\rm constant}).
$$
$$(18)$$  
where $J$ and $\tilde E$ are particle$^,$s angular momentum and energy per unit mass respectively.
The Newtonian approximation of the formula (18) is
$$ 
\frac{1}{2}\left (\frac {dr }{d t}\right )^2+\frac {J^2 }{2 r^2} -  \frac {GM}{r} –- \frac{c_0^2}{2}=E- \frac{c_0^2}{2} \approx \tilde E.
$$
$$(19)$$ 
Therefore, $\tilde E$ differs from $E$ in Newtonian gravity by a constant $c_0^2/2$.
  
The formulas (16), (17) and (18) are exactly the geodesic equations of EGR. According to EGR, spacetime is curved and all the coordinates $t, r, \phi $ in (13) do not have the direct meaning of time, distance, angle. In FGR, however, they do have, because spacetime is flat. Ironically, relativists when confronting GR to observational data, calculate time, distance, or angle by directly using the coordinates in (13).    
Therefore, all hailed accurate verification of general relativity is in fact the verification of FGR. 

Because $L\equiv H$, the upper limit of $\tilde E$, the energy per unit mass, is
$$ 
\tilde E_{{\rm max}} =0.
$$
$$(20)$$ 
For simplicity, I consider only the radial motion of particles with respect to the central mass. That is, they do not have angular momentum $J$, 
$$ 
J=0
$$
$$(21)$$
Choosing $J=0$ and $dr/d t=0$ in (18), we have the lower limit of energy $\tilde E$,
$$ 
\tilde E_{ {\rm min }}=-\frac{c_0^2}{2(1-2r_g/r)}. 
$$
$$(22)$$
Choosing $J=0$ and $\tilde E=0$ in (18) we have varying speed of radial light beam,
$$ 
c(r )=c_0\left (1-\frac{2r_g}{r}\right ), 
$$
$$(23)$$
Choosing $J=0$ and taking derivatives with $ t$ on the two sides of (18), we have the formula of radial acceleration for both light and material particles,
$$ 
a(r )=\frac {d^2r}{dt^2}= \frac{2r_g}{r^2} \left (1-\frac{2r_g}{r}\right ) 
\left (c_0^2+3\left (1-\frac{2r_g}{r}\right ) \tilde E \right ) .
$$
$$(24)$$
Zero energy ($ \tilde E =0$) corresponds to the motion of light.
Taking $\tilde E =0$ in (24), we have the radial acceleration of light,
$$ 
a_{ {\rm max }}=\frac {d^2r}{dt^2}= \frac{2r_g}{r^2} \left (1-\frac{2r_g}{r}\right ) 
c_0^2 >0. 
$$
$$(25)$$
Because the radial acceleration is positive, light decelerates toward the central mass. Therefore, light suffers repulsive force from the mass, contrary to people$^,$s imagination. This result of light deceleration is verified by the radar-echo-delay experiments (Shapiro, 1968 and 1971) and other similar experiments. 

Substituting the lower limit of energy (22) into (24), we have the lower acceleration limit for material particles 
$$ 
a_{ {\rm min }}=\frac {d^2r}{dt^2}=- \frac{r_g}{r^2} \left (1-\frac{2r_g}{r}\right )c_0^2 <0
$$
$$(26)$$
which is negative and corresponds to positive acceleration to the mass center. Therefore, low energy bodies do suffer attracting force (gravity!).

For the earth, $r_g=4.43 \times 10^{-3} $\,m. Near the earth surface, the formula of radial acceleration (24) can be approximated as
$$ 
a(r_e )= \frac {d^2r}{dt^2}=2g+\frac{6g}{c_0^2}\tilde E
$$
$$(27)$$
where $r_e$ is the earth radius and $g$ is the absolute value of the familiar acceleration
$$ 
g\approx 9.8 \;{\rm m\,s}^{-2}.
$$
$$(28)$$
From the formula (27) we know that material particles of radial motion suffer no gravity if their energy per unit mass reaches $-c_0^2/3$. That is,  particles$^,$ speed approaches light speed: $v\approx c/\sqrt{3}$.
From the formula (27), the minimum acceleration near earth surface is
$$ 
a_{ {\rm min }}  \approx -\frac {r_g}{r^2}c^2= -g 
$$
$$(29)$$
which is what Einstein thought to be the constant acceleration for all test particles of whatever energy near the earth surface. Einstein generalized this false thought as the equivalence principle and suggested the geometrization of gravity, the curved spacetime assumption. I return to the discussion in the next Section.

{\bf (vii) Flat-spacetime General Relativity (FGR). }
Special relativity describes the properties of global inertial frames in which the distribution of matter is both spatially and temporally homogeneous. Light speed is constant in all inertial frames and the formulation of physical laws is covariant with Lorentz transformations between the inertial frames. Both EGR and FGR introduce in-homogeneity into spacetime. EGR assumes curved spacetime and does not have global inertial frames. Its formulation is covariant with all possible curvilinear coordinate transformations. My FGR is based on flat spacetime. Is it true that the formulation of FGR is only covariant with all Lorentz transformations? The answer is no, which is explained in the following.

FGR maintains what is successfully testified in EGR. Einstein field equation which connects curvature tensor to matter, is a tensor equation and fits solar observation with great accuracy. The important example is the post Newtonian formulation of the equation. However, relativists when fitting the formulation to data, calculate time, distance, or angle by directly using the coordinates.
Therefore, relativists verified flat-spacetime according to Riemann theorem and the curvature tensor does not describe the curved spacetime at all. Therefore, FGR maintains all formal tensor calculus and keeps Einstein field equation.   
Because spacetime is flat, all tensors including the curvature one do not describe curved spacetime at all. For example, the covariant derivative
$$ 
\frac {\partial V^\mu}{\partial x^\lambda }+ \sum_\nu \Gamma ^\mu_{\lambda \nu}V^\nu 
$$
$$(30)$$
has no geometric meaning. This idea of flat-spacetime tensor calculus is not my invention. It is employed in the fluctuation theory of thermal physics many years ago.
Therefore, the answer to the above question is no and all physical law must be covariant with all coordinate transformations between real reference frames (generally they are freely falling frames).
If $x^\alpha , \;\alpha =0,1,2,3$ are rectangular coordinates on the flat spacetime then the following absolute derivatives
$$ 
\frac {\partial }{\partial x^\alpha }, \;\;\alpha =0,1,2,3 
$$
$$(31)$$ 
are not a covariant vector in FGR.

The covariance with all coordinate transformations between real reference frames provides the dynamical calculation of gravity in FGR. For example, all familiar global inertial frames are actually approximate ones. The static frame on earth which is considered to be inertial frame in civil building design is an approximate one with respect to the solar frame. Therefore, the rectangular coordinate in the earth frame is actually a curvilinear one in the solar frame. Therefore, covariant transformation of the curvilinear coordinates into the solar rectangular coordinates gives more accurate calculation of earth problem.     

{\bf (viii) Freely-Falling Frames in Flat-spacetime Gravity (FGR). }
The next Section shows that there is no local common acceleration for all test particles which cancels gravity as suggested by Einstein. In my flat-spacetime theory of gravity (FGR), the freely-falling frames with their coordinate axes being parallel transported according to the non-geometric connection, cancel gravity incompletely but mostly and globally. And the degree of cancellation differs with different freely-falling frames. Therefore, we people on earth frame can not feel the gravity from sun mostly. We can feel the gravity from the earth completely because we are not in the freely-falling frame with respect to earth$^,$s gravity.  

The next Section proves the existence of the unique inertial frame of the universe. Stars are in freely-falling frames with respect to galactic gravity, and the reference frames of stars can not detect much gravity from galaxies. Sun is such a star, and all our astronomic observation is based on the sun frame. {\bf That is why we see that all electro-magnetic waves from the universe demonstrate approximately the same physics!}\\
\\ 
{\bf FGR Provides a Consistent Explanation to Solar System and Galaxies and the Universe }\\
\\
Gravity is fundamental to the understanding of solar system, galaxies and the structure and evolution of the Universe. The curved-spacetime theory of gravity, EGR, encounters many difficulties in the explanation of large-scale systems 
(galaxies and the universe) and always resorts to dark matter and/or dark energy. Relativists claimed accurate tests of general relativity (GR) in the solar system. For example, the experiment of Gravity Probe B (GPB) claims the unprecedented accurate measurement and its result will be released within half a year. However, if GPB claims almost null error of GR prediction then the curved-spacetime assumption (EGR) is wrong and FGR is confirmed. This is because relativists when confronting GR to observational data, calculate time, distance, or angle by directly using the coordinates in Schwarzschild solution or in post Newtonian formulation or in the calculation of gravitational waves, and relativists are actually assuming flat spacetime according to Riemann$^,$s theorem. Therefore, these claims are specious to EGR.  
In the following, Einstein$^,$s equivalence principle is shown to be completely false and serves as the second specious claim to EGR. Then I review the simple and consistent FGR explanation to galaxies and the universe, and finally I show that the possibility is less than one in billion that the assumptions of curved spacetime, black holes, and the big bang are true.    

{\bf (i) Einstein$^,$s Equivalence Principle is False. }
Einstein$^,$s equivalence principle is that, over any small region of space and time, all test particles move at approximately the same acceleration. Therefore, the observational frame which moves at the very acceleration will see each particle being either static or moving in straight lines at constant speeds, within the small region in question. That is, the local frame sees no gravity at all and we see a cancellation of gravity by choosing local frames, which are generally called the local freely-falling frames. Einstein thought that the local frames were the local tangent 4-dimensional planes to the curved spacetime. This mistake led to the assumption of curved spacetime and resulted in ninety years$^,$ dogmatic study of gravity and cosmology: black holes, big bang, inflation, etc. I make two points to prove Einstein$^,$s mistake.

Firstly, a tangent plane is an inertial frame in which particles move in straight lines at constant speeds. Different particles may have different speeds but their acceleration must be zero. Speed is the first derivative of distance coordinate with time coordinate while acceleration is the second derivative. Tangent plane to curved space is determined only by the first derivatives not the second derivatives. How did Einstein make such simple mistake?

Secondly, Einstein made further mistake and assumed that all local particles shared the same acceleration independent of their individual properties, that is, independent of their energy per unit mass and their angular momentum per unit mass. Energy and angular momentum have totally four degrees of freedom and Einstein required that locally particles have zero degree of freedom: sharing the same acceleration.  
Based on FGR (see part (vi) of Section 1), however, I proved that particles must have different local accelerations corresponding to their angular momentum and energy. If angular momentum is zero (radial motion), the formula of acceleration is (24) which depends on energy. Only when their angular momentum is zero and their energy is the minimum will the test particles share the same acceleration. In the case of earth surface, the shared acceleration is $9.8$ m/s$^2$. If the energy of test particles were high enough then the leaning tower experiment of Galileo Galilei would have demonstrated opposite result. Einstein thought the result of Galileo Galilei were universal truth and generalized this false result as his equivalence principle.

Therefore, there is no local common acceleration which cancel gravity  locally, and Einstein$^,$s equivalence principle is completely false and serves as the second specious claim to EGR. However, a great many arguments of dogmatic gravitational theory and its applications and the theory of big bang are based on the non-existent freely-falling frames which cancel local gravity. 

In my flat-spacetime theory of gravity (FGR), the freely-falling frames with their coordinate axes being parallel transported according to the non-geometric connection, cancel gravity incompletely but mostly and globally. And the degree of cancellation differs with different freely-falling frames. Therefore, we people on earth frame can not feel the gravity from sun mostly. We can feel the gravity from the earth completely because we are not in the freely-falling frame with respect to earth$^,$s gravity.

{\bf (ii) EGR can not Explain Galactic Phenomena while FGR can. }
Solar system has the overwhelming mass point, the sun, and is generally considered to be a two-body problem. Einstein field equation can reduce to Newtonian gravity and proved to be useful in the two-body problem. However, Einstein field equation encounters many difficulties when applied to many-body systems like star clusters, galaxies, and the universe. Relativists try hard in such applications because they believe that spacetime is really curved and has the curvature tensor contained in Einstein field equation. Therefore, EGR assumes that all many-body systems of whatever shapes and scales, demonstrate the same attractive two-body phenomenon described by Newtonian gravity.
For example, EGR requires that galactic rotational curves be falling in radial direction from spiral galaxy centers. However, real observational data shows the opposite rising curves. In addition, EGR has no idea about why spiral galaxies have  2-dimensional disks. 

Because of modern powerful observational technique, galaxies have many established facts. The facts can be explained by my FGR. For example, the radial surface-brightness profile of spiral galaxy disks obeys exponential law. The spiral arms of spiral galaxies are curved waves in logarithmic curvature. 
FGR consistently explained the laws as well as the rising rotational curves (He, 2005a and 2005b). Based on FGR, stars in the 3-dimensional elliptical galaxies and in the 3-dimensional central bulges of spiral galaxies suffer attractive force towards their centers while the stars in the 2-dimensional disks of spiral galaxies suffer radial repulsive force.  
 
{\bf (iii) Flat-spacetime Model of the Universe Resolves the Difficulties Encountered by Big Bang Theory. }
We know that earth, sun, or a galaxy are all approximate inertial frames. Does the whole universe provide the unique accurate inertial frame? This frame is meaningful only if all galaxies slow down their motion and try to reach the final static spatial positions on the frame.
The existence of the unique inertial frame (e1: event 1) is proved based on my flat-spacetime model of the universe (He, 2006b). The model employs just one simple cosmological principle. The model not only explains galactic redshifts (e2) and Hubble redshift-distance law (e3) but also predicts a decreasing speed of light with time (e4) and the accelerating universe (e5). People have shown that decreasing speed of light resolves the difficulties encountered by big bang theory (BBT) (see Magueijo (2003)). The difficulties are horizon problem, flatness problem, etc. My model does not need dark matter and dark energy. Now I present the simple principle and my model (He, 2006b).
 
{\it Homogeneous yet evolving universe on flat spacetime (the cosmological principle).}
In the first half of last century, our knowledge of the universe was very limited and all models of the universe were mainly based on assumptions. Among the models was the big bang theory (BBT) which was based on curved spacetime assumption and became dogmatic. 
Now, cosmological study becomes an observational science and astronomical data does indicate that the large-scale universe is spatially homogeneous. That is, the universe is isotropic so that each observer sees the same in all directions. This is very strongly suggested by the observation of cosmic microwave background radiation (CBR): the temperature of CBR is independent of direction to one part in a thousand, according to a variety of experiments on various scales of angular resolution down to $1^\prime $ (Berry, 1989). 
However, BBT made many assumptions besides homogeneity which can not be observationally proved. The assumptions include bang-from-nothing, expansion, inflation, etc. To be fitted to data, more and more parameters were needed. When no more parameter can fix data, the un-observable stuff, dark matter and dark energy, was introduced. 

My flat-spacetime model is based on the single principle of spatial homogeneity which is observationally proved, and all above-said difficulties are gone. The Lagrangian which describes the motion of particles (galaxies, photons) in spatially homogeneous universe is unique as follows,  
$$ 
L\left(\tilde t, x^i, \frac{d\tilde t}{dp} , \frac{dx ^i}{dp}\right) = -\frac{1}{2} B(\tilde t ) \left(\frac{d\tilde t}{dp}\right)^2 
 + \frac{1}{2}A(\tilde t ) \left(\left(\frac{dx}{dp}\right)^2  + \left(\frac{dy}{dp}\right)^2 + \left(\frac{dz}{dp}\right)^2\right)
$$
$$(32)$$ 
where $\tilde t \equiv x^0 \equiv c_0t$, both $A(\tilde t) (>0)$ and  $B(\tilde t)(>0)$ depend on time $\tilde t$ only. If both $A(\tilde t)$ and  $B(\tilde t)$ are constants then the distribution of matter in the universe is also temporally homogeneous, no gravitational interaction is present on the cosmologic scale, and the Lagrangian simply returns to Einstein$^,$s special relativity. We assume that $A$ and $B$ vary with time and we have a spatially homogeneous yet evolving universe. This temporal inhomogeneity brings ``gravitational interaction$^{,,}$ into the components (galaxies and photons) of the universe. Because the universe is spatially homogeneous, the ``gravitational force$^{,,}$, at each spatial position, exerts in all spatial directions and the magnitude of the force is the same for all directions. Therefore, the ``gravity$^{,,}$ is called pressure gravity because it has the similar property to the one of pressure.  Einstein$^,$s equivalence principle definitely fails to the pressure gravity. 
  
The motion of particles (galaxies and photons) is the solution of the corresponding Lagrange$^,$s equation,
$$ 
\frac {dx^i}{d\tilde  t}= -P_i \sqrt{ \frac{ B(\tilde  t)}{(P^2-2\tilde EA(\tilde  t)) A(\tilde  t)}}. 
$$
$$(33)$$  
where $P_i, i=1,2,3$ are the conservative spatial momentum vector with $P$ being its amplitude. However, this is not the full story. Since we deal with causal motion only, we always have $\tilde E \leq 0$. 

{\it Varying light speed in the gravitational field of the universe. } 
Because light has the maximum speed, we have $\tilde E = 0$ for the motion of light. In its propagation direction we have
$$ 
\frac {dx}{d\tilde  t}= \sqrt{ \frac{ B(\tilde  t)}{A(\tilde  t)}}. 
$$
$$(34)$$  
Currently the universe is at the time of
$$ 
\tilde t=\tilde t_1 =ct_1.
$$
$$(35)$$  
The current light speed is $c \simeq 3\times 10^8 {\rm m\,s}^{-1}$ which is used in the definition of $\tilde t $: $\tilde t = ct$. It is not wrong that we choose other light speed for the definition.   

{\it Galactic redshift and Hubble law. } 
Galactic redshift is the formula (48) in He (2006b),
$$ 
z=\frac{\nu _1 }{\nu _2} -1= \frac{ \sqrt{g _{00}( \tilde  t _1)} }{\sqrt{g _{00}( \tilde  t _2)}}-1 
=\frac{ \sqrt{ B(\tilde  t_1 ) } }{\sqrt{B(\tilde  t _2) }} -1.
$$
We see that $B(\tilde  t )$ must be a monotonously increasing function with time for us to have galactic redshifts rather than blueshifts,
$$ B(\tilde t) \uparrow . $$  
The distance $D$ between the two galaxies 1 (Milky Way) and 2 is given by the integral of the light travel formula (34)
$$ 
D= \int ^{\tilde  t_1} _{\tilde  t _2}\frac {dx}{d\tilde  t}d\tilde  t= \int ^{\tilde  t_1} _{\tilde  t _2} \sqrt{ \frac{ B(\tilde  t)}{A(\tilde  t)}} d\tilde  t. 
$$ 
The distance formula must have a redshift factor to give the Hubble law.
This indicates that $A(\tilde  t )$ depends on $B(\tilde  t )$. A simple and general model of the dependence is
$$ 
A(\tilde  t )= \frac {B^{m+1}(\tilde  t)}{N^2B^{\prime 2} (\tilde  t) }
$$
$$(36)$$ 
where $m$ is a constant and $N(>0)$ is another constant whose unit is length.
Finally we have Hubble law,
$$
\begin{array}{ll}
D&= \frac{2N}{m-2}\left (\frac{1}{\sqrt{ B(\tilde  t_2) }^{m-2} } -\frac{1}{\sqrt{ B(\tilde  t_1) }^{m-2} }\right )\\
&= \frac{2N}{m-2}\left (\frac{1}{\sqrt{ B(\tilde  t_2) } } -\frac{1}{\sqrt{ B(\tilde  t_1) } }\right )\left(\frac{1}{\sqrt{ B(\tilde  t_2) }^{m-3} } +\dots  \right ) \\
&=\frac{2Nz}{(m-2)\sqrt{ B(\tilde  t_1) }}\left(\frac{1}{\sqrt{ B(\tilde  t_2) }^{m-3} } +\dots  \right ) \\
&=\frac{cz}{H_0(\tilde  t_2, \tilde  t_1)} 
\end{array}
$$
$$(37)$$   
where the Hubble constant $H_0$ is
$$ 
H_0=\frac{c(m-2)\sqrt{ B(\tilde  t_1) }}{2N}/\left(\frac{1}{\sqrt{ B(\tilde  t_2) }^{m-3} } +\dots  \right ). 
$$
$$(38)$$ 
As a summary, I note that the redshift requires $B(\tilde t )$ be a monotonously increasing function of time and Hubble law requires $A$ be determined by the function $B$ (see (51)).  
Therefore, the only one degree of freedom left is the function form of $B(\tilde t )$. 

{\it 'Accelerated Expanding$^,$ Universe.}
If $H_0$ depended only on $\tilde  t_1$, the current time, then Hubble law would be perfectly true.  However, it depends on the past time of the galaxy we observe,  
$$ 
H_0= H_0(\tilde  t_2, \tilde  t_1).
$$
$$(39)$$ 
If we assume
$$ 
m>3
$$ 
then Hubble constant $H_0$ is not constant and increases with the time $\tilde  t_2$, of which the galaxy is observed.  This increase with time of $H_0$ is explained as the `accelerating expansion$^,$ of the universe. However, in my model, spacetime is flat (no expansion of curved spacetime) and the redshift is gravitational one which results from the evolution of the universe (mass density varies with time).
Because redshift requires increasing $B(\tilde  t)$, we see that `accelerating expansion$^,$ is consistent to galactic redshift.
    
{\it Infinite Light Speed and the Birth of the Universe.}  
Positive and increasing quantity $B(\tilde  t)$ indicates a time $\tilde  t _0$,  when $B(\tilde  t _0)=0$. This is the starting time of the universe. We can choose $\tilde  t _0= 0$ to be the time of birth. Currently we do not know the exact physics at the hot birth. One thing is sure that light speed at the time must be infinite. Only infinite speed of communication could result in a later spatially homogeneous mass distribution in the infinite flat universe. This resolves the horizon and flatness problems due to birth. Infinite initial light speed indicates a decrease of light speed with time. Observation during the last decade does support the result of decreasing light-speed with time. The formula of light speed is (34). Therefore, decreasing light speed imposes further condition on the evolving factor $B(\tilde  t)$,
$$ 
     2BB^{\prime \prime} \le m{B^\prime}^2.
$$
$$(40)$$

{\it Light Speed Constancy and the Death of the Universe.}
However, there is strong evidence that light speed is approximately constant during mature stage of the universe. Constant light speed with time means that $A(\tilde  t) $ and $B(\tilde  t)$ are the same 
$$ 
  A(\tilde  t) \equiv B(\tilde  t) .
$$ 
They serve as the scaling factor. Perfect Hubble redshift-distance linear law completely determines the scaling factor,
$$ 
     \frac {1} {B(\tilde  t) } \equiv \frac {1} {A(\tilde  t) } = \frac {1} {B_0 } –- M (\tilde t - \tilde  t_0) 
$$
$$(41)$$ 
where $M$ is a constant and $ B_0 = B(\tilde  t_0) $. This formula indicates a finite time $ \tilde  t_1$  when $ M(\tilde  t_1-\tilde  t_0  )= 1/B_0 $.  This is the ending time of the universe because the scaling factor reaches infinity. The possibility of a rebirth needs further investigation.  

{\it The Absolute Inertial Frame of the Universe.}
Our calculation and results are reference-frames depended. For example, photon frequency is dependent on reference frames. Our results are meaningful only when single preferred inertial frame of the universe exists and the results are calculated with respect to the frame. The absolute frame is meaningful only when all components (e.g., galaxies) of the universe have convergent motion with respect to the frame. That is, all components slow down their speed of motion with respect to the frame. Since the nineteenth century, scientific report on the evidences of absolute inertial frame has never been stopped. Because of light speed constancy we have $  A(\tilde  t) \equiv B(\tilde  t) $ in the formula (33). We can see that the absolute speed of material particles (galaxies) does decrease with time, slowing-down motion with respect to the absolute inertial frame (note that $\tilde E < 0$ for material particles). Here we see that the existence of absolute inertial frame is once again the direct result of galactic redshift.  

{\it The Variance with Time of Matter Distribution in the Universe.}

Our Lagrangian is defined on flat spacetime and can be quantized according to the classical and covariant quantization procedure (He, 2006a). Because the spatial distribution of matter in the universe is homogeneous, the resulting amplitude of the wave function must be proportional to the density of the distribution. Astronomical observation suggests that the density decreases with time especially during early universe. We can see that the amplitude does decrease with time if $B(\tilde  t) $ increases with time. That is, the astronomic observation is once again consistent to the result of galactic redshift.

{\bf (iv) The Possibility of Curved Spacetime, Black Holes, and Big Bang is Less than One in Billion. }
You have probably noticed that my FGR is based on very simple principles. Now I calculate the probability that FGR is wrong. FGR generalizes special relativity (e6, event 6). Einstein$^,$s general relativity does not generalize SR (special relativity). Because SR is well verified in high energy physics, the probability is less than one in hundred ($10^{-2}$) that the requirement of a gravitational theory which generalizes SR is false. My FGR explains the phenomena of galaxies (e7), which is false with the possibility of less than a hundredth ($10^{-2}$).   
My FGR quantizes gravity (e8), which is false with the possibility of less than a hundredth ($10^{-2}$).  My model of the universe predicts many observational facts. Its single principle is that the universe is evolving (e9).       
The probability is less than one in hundred ($10^{-2}$) that the universe is not evolving (e9 is false). My model of the universe involves the single function of time: $B(\tilde t)$. The function is arbitrary except satisfying some conditions. Redshifts require increasing $B(\tilde t)$ with time. Decreasing speed of light requires that $B(\tilde t)$ satisfies the condition (40). The simple conditions guarantee the existence of the unique inertial frame of the universe (e1), the redshifts (not blueshifts) of galaxies (e2), the Hubble redshift-distance law (e3), the decreasing light-speed which resolves big bang difficulties (e4), and the `accelerating expansion` universe (e5). These predictions of independent observational cosmological facts based on the two conditions of single arbitrary function are certainly not an accident. Therefore, the probability that my model of the universe is not scientific truth is less than one in million ($10^{-6}$). Because these observational facts and the principles are independent events, the probability that FGR is false is less than one in billion ($10^{-9}$):
$$10^{-6} \times 10^{-2} \times 10^{-2} \times \cdots < 10^{-9}.$$
That is, the possibility that the assumptions of curved spacetime, black holes, and big bang are true, is less than one in billion.

{\bf (v) Why we See that All Electro-magnetic Waves from the Universe Demonstrate Approximately the Same Physics. }
I have shown the existence of the unique inertial frame of the universe. Stars are in freely-falling frames with respect to galactic gravity, and the reference frames of stars can not detect much gravity from galaxies. Sun is such a star, and all our astronomic observation is based on the sun frame. {\bf That is why we see that all electro-magnetic waves from the universe demonstrate approximately the same physics!}\\
\\
{\bf Suggesting an Experiment to Test Curved-Spacetime Assumption  }\\
\\
{\bf (i) ``No Anysotropy of Light Speed is Observed$^{,,}$: the Third Specious Claim to EGR. }
Experimentalists of relativity claim that no anisotropy of light speed is observed. That is, no evidence of different light speeds is found. That means that light in vacuum demonstrates the unique value independent of its origin and reference frame. If their claim were correct then EGR must be wrong. This is because EGR is the theory of gravity and gravity introduces spatial or temporal in-homogeneity. 
Only if the distribution of matter were both spatially and temporally homogeneous could we have a global inertial frame where Einstein$^,$s special relativity would be perfectly true and light speed would be constant for all inertial frames. If such homogeneity does happen and SR is perfectly true in the inertial frame, light speed is definitely anisotropic in any non-inertial frame.
 
Ironically, all experiments measuring light-speed anisotropy were performed on earth. The rotating earth is neither an inertial frame in FGR nor a freely falling frame in EGR.
According to EGR, light speed is constant only in the local inertial frames (the local tangent ``planes$^{,,}$ to curved spacetime). There is no such stuff as local freely-falling frames which cancel gravity (see part (vi) of Section 1). Therefore, light speed in rotating earth frame is definitely anisotropic according to EGR (I look forward to some relativist who will derive the anisotropy formula of light-speed in non-inertial frames as predicted by EGR). 
Light-speed anisotropy in rotational frames was proved by Sagnac experiment and relativists admitted that light speed is not constant in non-inertial frames. 
Because experimental relativists claimed no measurement of light-speed anisotropy on earth frame which is against  theoretical relativists$^,$ expectation, EGR is wrong. Therefore, relativists made the third specious claim to EGR.   
  
{\bf (ii) Anisotropy Formula of Light-Speed Based on FGR. }
According to FGR which is based on flat spacetime, gravity results in varying light speed in inertial frames. If the pattern of varying light-speed is measured to conform to the formula of FGR then the curved-spacetime assumption is dead. Now I derive the anisotropy formula of light-speed based on FGR. 
 
In Section 1, I calculated the speed of radial light beam which is (23). Choosing $\tilde E=0$ and $dr/d t=0$ in (18), we have the speed of light beam in the perpendicular direction to the radial one,
$$ 
c(\pi/2)=r\frac{d\phi }{dt}=c_0\sqrt{1-2r_g/r }. 
$$
$$(42)$$ 
The corresponding angular momentum $J$ is the maximum,
$$ 
J_{ {\rm max}}=\frac{rc_0 }{\sqrt{1-2r_g/r }}. 
$$
$$(43)$$ 
If light beam makes the angle $\theta $ with respect to the radial direction then its angular momentum is between zero and the maximum
$$ 
J=f\frac{rc_0 }{\sqrt{1-2r_g/r }}, \;\; 0\le f\le 1. 
$$
$$(44)$$ 
It is straightforward to show that 
$$ 
f\approx \sin \theta .
$$
$$(45)$$ 
The formula of light-speed anisotropy is
$$  
c^2(\theta )  =  c_0^2( 1-2r_g/r)^2\left(1+f^2 \frac {2r_g/r}{1-2r_g/r}\right).
$$
$$(46)$$ 
Therefore, 
$$  
c(\theta ) \approx c_0\left(1+\frac{r_g}{r}\sin^2\theta -\frac {2r_g}{r} \right)
$$
$$(47)$$ 
which is the required anisotropy formula of light-speed in FGR.

The formula (47) is the speed of light beam staring at the radial position $r$ and running in the direction which makes a angle $\theta $ to the radial direction. No modern technique can measure the speed of single light beam to the accuracy of about kilometers per second. However, modern technique can measure  the difference of light speeds of two light beams to the accuracy of about $10^{-6}$ meters per second. Therefore, the experiments for measuring anisotropy generally have two light beams starting at the same position $r$. One-way experiment lets them travel a small distance in their different directions and then measure their light-speed difference.  Two-way experiment requires them return to their starting position and then measure their light-speed difference. From the formula (47) we know that the light-speed difference is the maximum ($\Delta c \approx r_g/r$) only when one beam runs in the radial direction with respect to the mass center and the other runs in the perpendicular direction. 

Now let us study such experiment on earth surface. We have already known that the maximum magnitude of anisotropy is $\Delta c \approx r_g/r$ where $r$ is the radial distance to the mass center. The experiment on earth surface involves two mass centers. One is the sun and the other is the earth. It is interesting that the magnitude of anisotropy due to sun is $\Delta c \approx 3$ m/s which is about ten times larger than the one due to earth, $\Delta c \approx 2\times 10^{-1}$ m/s. 
However, earth is the freely-falling frame with respect to the sun whose gravitational effect can not, mostly, be detected from the experiment on earth. 
Therefore, if FGR is true then the measured anisotropy on earth surface must be due to earth. However, rotating earth is non-inertial frame. Because people believe that the anisotropy of light-speed due to earth$^,$s rotation is $\Delta c \approx c_0(v_e/c_0)^2$ ($v_e$ is the linear velocity of earth rotation at equator) which is even smaller than the anisotropy due to the mass of earth (see Klauber, 2006), the effect of earth rotation is neglected. Therefore, our goal is to test anisotropy of light-speed on earth due to the gravity of earth. According to the analysis given in the last paragraph, to achieve the maximum magnitude of light-speed difference we need to direct one light beam to the mass center of earth and the other beam in the perpendicular direction, i.\,e., the parallel direction to the earth surface. I forgot the final condition: the experiment has the ability to measure light-speed difference to an accuracy better than $ 0.2$ m/s.                 

{\bf (iii) Suggesting an Experiment to Test Curved-Spacetime Assumption. }
The only experiments which claimed the above accuracy were performed by Brillet and Hall (1979), Hils and Hall (1990), and Muller {\it et al.} (2003). However, all the experiments aimed at testing Einstein$^,$s special relativity. That is, they test the anisotropy of light-speed due to absolute motion with respect to the absolute reference frame, the aether. I call it aether-frame anisotropy. My FGR is based on the assumption that special relativity is correct and provides the anisotropy formula of light speed (47) due to central gravity of mass point. 

The formula of aether-frame anisotropy is dogmatically derived and is given as follows.
If at $t=0$ a beam of light is emitted in $\Sigma $  and if $S$ 
(non-preferential frame) moves with the speed $v$ with respect to $\Sigma $  and if $v$ makes the angle $\theta $  with respect to the direction of the light beam then $S$ measures for light the speed $c$, where
$$  
c(\theta , v)  =  c_0\left(1 + \frac{ Av^2}{c^2} \sin^2\theta  +  \frac{ Bv^2}{c^2}\right)
$$
$$(48)$$ 
Parameter $A$ is a measure of light speed isotropy and is generally measured through a 
Michelson-Morley class of experiments.  These experiments verify light speed isotropy. 
Parameter $A$ has been tested by to be less than $3\times 10^{-15}$.
Parameter $B$ is a measure of light speed invariance relative to the speed of the emitter/receiver and it is generally measured through Kennedy-Thorndike experiments. 
These experiments verify light speed invariance with the movement of the 
emitter/observer.  Parameter $B$ has been tested by to be less than $3\times 10^{-13}$.  SR predicts $A=B=0$ and the experimental asymptotical limits for both $A$ and $B$ under SR are indeed zero. 

However, all experiments were designed to test aether-frame anisotropy. Therefore, they do not satisfy the conditions required to testify the anisotropy due to central gravity. For example, we consider the Brillet and Hall experiment. Firstly, the two light beams were both parallel to earth surface. 
Therefore, it is not surprised that it gives null result of light-speed anisotropy. Secondly, they were forced to subtract out a ``spurious$^{,,}$ and
persistent signal of approximate amplitude $2\times10^{-13}$
at twice the rotation frequency of their apparatus (Klauber, 1999). Thirdly, for the purpose of fitting data to the formula of aether-frame anisotropy (48), they averaged out some daily periodic variation and subtract away some seasoned pattern. The formula (47) actually predicts just such an effect due to the central mass of sun and the earth rotation.     

Therefore, I suggest to repeat these experiments with one light beam in gravity direction so as to test the anisotropic light-speed due to earth gravity.
Dare to make public the recent result of new Brillet and Hall experiment with one vertical light beam?   \\
\\
{\bf Conclusion  }\\
\\
{\bf (i) The First Specious Claim Made for EGR.}
Einstein$^,$s general relativity (EGR) is the theory of curved spacetime. 
However, his assumption brings more complexity than truth.
Firstly, curved spacetime is embodied by non-trivial topology. Because topology is a very complicated mathematical subject, most relativists never take a look at it. Secondly, the concept of curved spacetime is nothing but temporal and spatial in-homogeneity. Therefore, all coordinates on a curved space are merely parameters. Real time and distance have to be calculated by employing  coefficients of the spacetime metric. The calculation of time and distance by employing metric is very complicated too. 
Therefore, all relativists when confronting GR to observational data, calculate time, distance, or angle by directly using the coordinates in Schwarzschild solution or in post Newtonian formulation. 
However, there is the famous Riemann theorem: 
only when spacetime is flat does there exist one coordinate system which has
direct meaning of time and distance, and {\it vice verse}.
Therefore, the hailed accurate tests of GR verified the flat-spacetime interpretation of GR (my FGR). 
The more claims are made that classical tests of general relativity fits data with great accuracy, the more falsified is the curved-spacetime assumption. That is, the claim is specious to EGR.

{\bf (ii) The Second Specious Claim Made for EGR.}
Einstein$^,$s equivalence principle is that, over any small region of space and time, all test particles move at approximately the same acceleration. Therefore, the observational frame which moves at the very acceleration will see each particle being either static or moving in straight lines at constant speeds, within the small region in question. That is, the local frame sees no gravity at all and we see a cancellation of gravity by choosing local frames, which are generally called the local freely-falling frames. Einstein thought that the local frames were the local tangent 4-dimensional planes of curved spacetime. This mistake led to the assumption of curved spacetime and resulted in ninety years$^,$ dogmatic study of gravity and cosmology: black holes, big bang, inflation, etc. I have made two points to prove Einstein$^,$s mistake.

Firstly, a tangent plane is an inertial frame in which particles move in straight lines at constant speeds. Different particles may have different speeds but their acceleration must be zero. Speed is the first derivative with particles$^,$ coordinates while acceleration is the second derivative. Tangent plane to curved space is determined only by the first derivatives not the second derivatives. How did Einstein make such simple mistake?

Secondly, Einstein made further mistake and assumed that all local particles shared the same acceleration independent of their individual properties, that is, independent of their energy per unit mass and their angular momentum per unit mass. Energy and angular momentum have totally four degrees of freedom and Einstein required that locally particles have zero degree of freedom: sharing the same acceleration.  
Based on FGR (see part (vi) of Section 1), however, I proved that particles must have different local accelerations corresponding to their angular momentum and energy. If angular momentum is zero (radial motion), the formula of acceleration is (24) which depends on energy. Only when their angular momentum is zero and their energy is the minimum will the test particles share the same acceleration. In the case of earth surface, the shared acceleration is $9.8$ m/s$^2$. If the energy of test particles were high enough then the leaning tower experiment of Galileo Galilei would have demonstrated opposite result. Einstein thought the result of Galileo Galilei were universal truth and generalized this false result as his equivalence principle.

Therefore, there is no such stuff as freely-falling frames which cancel local gravity, and Einstein$^,$s equivalence principle is completely false and serves as the second specious claim to EGR. However, a great many arguments of dogmatic gravitational theory and its applications and the theory of big bang are based on the non-existent freely-falling frames which cancel local gravity.

{\bf (iii) The Third Specious Claim Made for EGR.}
Experimentalists of relativity claim that no anisotropy of light speed is observed. That is, no evidence of different light speeds is found. That means that light in vacuum demonstrates the unique value independent of its origin and reference frame. If their claim were correct then EGR must be wrong. This is because EGR is the theory of gravity and gravity introduces spatial or temporal in-homogeneity. 
Only if the distribution of matter were both spatially and temporally homogeneous could we have a global inertial frame where Einstein$^,$s special relativity would be perfectly true and light speed would be constant for all inertial frames. If such homogeneity does happen and SR is perfectly true in the inertial frame then light speed is definitely anisotropic in any non-inertial frame.
 
Ironically, all experiments measuring light-speed anisotropy were performed on earth. The rotating earth is neither an inertial frame in FGR nor a freely falling frame in EGR.
According to EGR, light speed is constant only in the local inertial frames (the local tangent ``planes$^{,,}$ to curved spacetime). There is no such stuff as local freely-falling frames (see part (vi) of Section 1). Therefore, light speed in rotating earth frame is definitely anisotropic according to EGR (I look forward to some relativist who will derive the anisotropy formula of light-speed in non-inertial frames as predicted by EGR). 
Light-speed anisotropy in rotational frames was proved by Sagnac experiment and relativists admitted that light speed is not constant in non-inertial frames. 
Because experimental relativists claimed no measurement of light-speed anisotropy on earth frame which is against theoretical relativists$^,$ expectation, EGR is wrong. Therefore, relativists made the third specious claim to EGR.   

However, my FGR has no such contradictory claims. When confronted to solar observation, to future GPB data, and even to the gravitational radiation
damping data in a binary pulsar system (e.\,g., PSR 1913+16), it is directly verified without the panic of directly using coordinates as time, distance, or angle. EGR has no idea about galaxies while my FGR solves galaxy pattern and dynamics completely (He, 2005a, 2005b, 2005c, 2007). Consistent to FGR, my model of the universe proved the existence of the unique global inertial frame. What is more important, it is very simple and gives simple explanation to all available laws of cosmological observation.  It is more consistent than Big Bang Theory (BBT). Because I have traditional flat spacetime, gravity is easily quantized (He, 2006a). 

EGR and FGR are the mirrored versions of each other. If they are the only choice towards the truth of gravity then one must be real and the other is its illusory, tortuous, specious image. However, I have shown that the possibility  of curved spacetime, black holes, and big bang, is less than one in billion.
An experiment is proposed whose results will completely decide the fate of curved spacetime assumption. Dare to make public the recent result of new Brillet and Hall experiment with one vertical light beam?  \\
\\
Berry, M.: 1989, {\it Principles of Cosmology and Gravitation}, Institute Of Physics Publishing, Bristol and Philadelphia
\\  Brillet, A. and Hall, J.\,L.: 1979, {\it Phys. Rev. Lett. }  {\bf 42}, 549
\\ Crothers, S.\,J.: 2005, {\it Progress in Physics }, {\bf 3}, 41
\\ He, J.: 2005a, http://arxiv.org/abs/astro-ph/0510535
\\ He, J.: 2005b, http://arxiv.org/abs/astro-ph/0510536
\\ He, J.: 2005c, http://arxiv.org/abs/astro-ph/0512614v3
\\ He, J.: 2006a, http://arxiv.org/abs/astro-ph/0604084
\\ He, J.: 2006b, http://arxiv.org/abs/astro-ph/0605213
\\ He, J.: 2007, Astrophy. \&Space Sci., accepted
\\ Hils, D. and Hall, J.\,L.: 1990, {\it Phys. Rev. Lett.} {\bf 64}, 1697
\\ Klauber, R.\,D.: 1999, {\it Am. J. Phys.} {\bf 67}, 158
\\ Klauber, R.\,D.: 2006, http://arxiv.org/abs/gr-qc/0604118
\\ Magueijo, J.: 2003, {\it Rept. Prog. Phys.}, {\bf 66}, 2025
\\ Muller, H. {\it et al.}: 2003, {\it Phys. Rev. Lett.} {\bf 91}, 020401-1
\\ Shapiro, I.\,I. {\it et al.}: 1968, {\it Phys.\,Rev.\,Lett.}, {\bf 20}, 1265
\\ Shapiro, I.\,I. {\it et al.}: 1971, {\it Phys.\,Rev.\,Lett.}, {\bf 26}, 1132

\newpage
\pagenumbering{arabic}

\begin{center}
\large{
Einstein`s Geometrization vs. Holonomic Cancellation of Gravity via Spatial Coordinate-rescale and Nonholonomic Cancellation via Spacetime Boost}
\normalsize\\
\mbox{   }     
\end{center}

\large {{\bf Abstract} } \normalsize 
Particle`s acceleration in static homogeneous gravitational field is cancelled by any reference frame of the same accelerating direction and the same accelerating rate. The frame is commonly called the freely-falling one. The present paper shows that the acceleration is also cancelled by a spatial curvilinear coordinate system. The coordinate system is simply a spatial square-root coordinate rescale in the field direction, no relative motion being involved. This suggests a new equivalence principle. Spacetime is flat which has inertial frame of Minkowski metric $\eta _{ij}$. Gravity is a tensor $g _{\alpha \beta }$ on the spacetime, which is called effective metric. The effective metric emerges from the coordinate transformation. The gravitational field of an isolated point mass requires a nonholonomic spacetime boost transformation. 
This generalization of Newtonian gravity shares the properties of Lorentz transformation, which should help quantize gravity.
The corresponding effective metric is different from that of Schwarzschild. To first order, its prediction on the deflection of light and the precession of the perihelia of planetary orbits is the same as the one of general relativity (GR). Its further implication is left for future work.
\\
Relativity -- Gravitational Theory -- Galaxies : Structure 
\\
\\
\section{ Introduction }

{\bf (i) Minkowski metric description of vanishing gravity.}
The present paper deals with gravitational interaction only, no other interaction being involved.
Newton`s first law of motion that a particle experiencing no net force (i.\,e., vanishing gravitational field) must move in straight direction with a constant (or zero) velocity with respect to  inertial frame $\tau \xi \eta \zeta $, can be proved geometrically by introducing Minkowski metric $\eta _{\alpha \beta }$ to the frame,
\begin{equation} 
\begin{array}{ll}
ds^2 &=d\tilde \tau ^2-d\xi ^2 -d\eta ^2 -d\zeta ^2 \\
     & =-\eta _{\alpha \beta }d\xi ^\alpha d\xi ^\beta
\end{array}
\end{equation}
where $\xi ^0=c\tau =\tilde \tau, \xi ^1=\xi ,\xi ^2 =\eta , \xi ^3 =\zeta $, $c$ is light speed, and $\eta _{00}=-1, \eta _{11}= \eta _{22} = \eta _{33} =1, 
\eta _{\alpha \beta } = 0 (\alpha \not= \beta) $. 
The metric is the basis of special relativity. I call the distance $s$ along the curves of spacetime by real distance because I will introduce a new term, effective distance $\bar s$. The real distance is generally called proper distance
which can be negative because the matrix $\eta _{\alpha \beta }$ is indefinite.  The indefinite quadratic form (1) is the generalization of Pythagoras theorem to Minkowski spactime. 
It is straightforward to show that the first Newton law of motion (vanishing gravity) is equivalent to the following geodesic equation, 
\begin{equation} 
\frac {d^2\xi ^\alpha }{dp^2}+\Gamma ^\alpha_{\beta \gamma }
\frac {d\xi ^\beta }{dp}\frac {d\xi ^\gamma }{dp}=0  
\end{equation}
where $p$ is the geodesic-curve parameter and $\Gamma ^\alpha_{\beta \gamma }$ is the affine connection. The affine connection involves the first order derivatives to $\eta _{\alpha \beta }$ and must be zero. Therefore,
the first Newton law of motion is equivalent to (2).
People try to generalize the equation to describe gravitational interaction.

{\bf (ii) Part-one assumption of general relativity.}
It is more important to consider test particle`s motion in an inertial frame in which the particle does experience gravitational force. In the frame, the particle no longer moves in straight direction with a constant (or zero) velocity. The motion is described in good approximation by the Newton`s universal law of gravitation which is, however, a non-relativistic theory and needs to be generalized to give account for the solar observations which deviate from Newton laws` calculation. Einstein`s general relativity (GR) is the most important try toward the generalization. The basic assumption of GR can break into two parts. The 
part-one assumption of GR is the simple replacement of the above matrix $\eta _{\alpha \beta }$ by a tensor field $g _{\alpha \beta }$ whose components are, instead of the constants $\pm 1$,  position functions on spacetime. Similar to the above part (i) description, particles` motion follows the solution of the geodesic equation
\begin{equation} 
\frac {d^2x ^\alpha }{dp^2}+\Gamma ^\alpha_{\beta \gamma }
\frac {dx ^\beta }{dp}\frac {dx ^\gamma }{dp}=0  
\end{equation}
where $x ^0=ct =\tilde t, x ^1=x ,x ^2 =y , x ^3 =z$ 
and the affine connection $\Gamma ^\alpha_{\beta \gamma }$ involves the first order derivatives to the tensor,
\begin{equation} 
\Gamma ^\alpha _{\beta \gamma }=\frac{1}{2}g^{\alpha \rho }\left (
\frac {\partial g_{\rho \beta }}{\partial x^\gamma } +
\frac {\partial g_{\rho \gamma }}{\partial x^\beta } -
\frac {\partial g_{\beta \gamma }}{\partial x^\rho } \right ).  
\end{equation}  
The equation (3) does not involve the inertial mass of the test particles. This is an appropriate description because inertial mass equals gravitational mass in the case of gravitational interaction.  

{\bf (iii) Part-two assumption of GR (geometrization).} 
The present paper questions the part-two assumption of GR. The assumption is that spacetime is curved when gravity is present and $g_{\alpha \beta }$ in (4) is exactly the metric of the curved spacetime
\begin{equation}  
ds^2 =-g _{\alpha \beta }dx ^\alpha dx ^\beta .
\end{equation}
The assumption is called the geometrization of gravity whose whole meaning is that $s$ must be the real distance along the curves of spacetime. Therefore, the solutions of the geodesic equation (3) extremize the following functional variation  
\begin{equation}  
\delta s=\delta \int ^{p_B}_{p_A}\frac {ds}{dp}dp .
\end{equation}

The geometrization is claimed to be based on the following simple fact on static homogenous gravity.
Particle`s acceleration in static homogeneous gravitational field is cancelled by any reference frame of the same accelerating direction and the same accelerating rate. This is straightforward because any test particle in the field, with fixed directions of its attached axes, sees other test particles moving on straight lines with constant speeds. The frames are commonly called the freely-falling ones, which are the exclusive property of homogeneous gravity. 
However, the simple fact is not the full story of homogeneous gravity. In the following part (v) I will show that the acceleration is also cancelled by a spatial curvilinear coordinate system. The coordinate system is simply a spatial square-root coordinate rescale, no relative motion being involved. 
Firstly, in the part (iv) which deals with freely-falling frames, I will show that geometrization does not apply to homogeneous gravity.  

{\bf (iv) Failure of the geometrization of homogeneous gravity. }Static homogeneous gravitational field, $\vec g$, in the positive direction of $x$-axis can be canceled by a global space-time coordinate transformation,
\begin{equation} 
\begin{array}{l}
\xi=x- \frac {1}{2} g t ^2  , \\
\tau =t.  
\end{array}
\end{equation}
where $g (> 0)$ is constant. That is, in the $\tau \xi $ coordinate system, particles experience no gravity and follow the equation (2) in part (i) where $\Gamma ^\alpha_{\beta \gamma }$ involves the first order derivatives to $\eta _{\alpha \beta }$ and must be zero. The real distance $\bar s$ in the freely-falling $\tau \xi$ frame is the formula (1),
\begin{equation} 
d\bar s^2 =d\tilde \tau ^2-d\xi ^2 
\end{equation}
where I introduced a new symbol $\bar s$ instead of $s$. Its explanation will be given in the following. 
For simplicity, I drop off the coordinates $\eta , \zeta , y, z$ when dealing with homogeneous gravity. 
Substitution of the formula (7) into (8) leads to a quadratic form in the coordinates $t, x$,
\begin{equation} 
\begin{array}{ll}
d\bar s^2 &=\left (1- \frac{g^2\tilde t ^2}{c^4}\right )d\tilde t^2
                +\frac{2g\tilde t}{c^2}d\tilde t dx - dx ^2   \\
     &=-g _{\alpha \beta }dx ^\alpha dx ^\beta
\end{array}
\end{equation}
where
\begin{equation} 
g_{00}=\frac{g^2\tilde t ^2}{c^4}-1, g_{01}=g_{10}=-\frac{g\tilde t}{c^2}, g_{11}=1.
\end{equation}
Now I apply the part-one assumption of GR, i.\,e. the method in part (ii), to the above quantity $g _{\alpha \beta }$. It is not surprising that   
the solution of the corresponding geodesic equation turns out to be $x=(1/2)gt^2 +x_1$ where $x_1$ is a constant, i.\,e.,
\begin{equation} 
\frac {d^2x}{dt^2}=g.
\end{equation}
This indicates that particles in $tx$ coordinate system experience static homogeneous gravitational field $\vec g$ and the coordinate transformation (7) does cancel gravity. 
However, the geometric explanation of $\bar s$ to be real distance on the spacetime fails, as demonstrated in the following.  
In fact, Einstein`s geometrization of gravity refuses any cancellation of gravitational field by a global spacetime coordinate transformation, because of a mathematical theorem. The theorem is that if the spacetime $txyz$ is curved then there is no global coordinate transformation 
\begin{equation} 
\begin{array}{ll}
t=t(\tau ,\xi ,\eta ,\zeta ), 
& x=x(\tau ,\xi ,\eta ,\zeta ), \\
y=y(\tau ,\xi ,\eta ,\zeta ), 
& z=z(\tau ,\xi ,\eta ,\zeta )
\end{array}
\end{equation}
which transforms the quadratic form (5) into (1), and, if there is such coordinate transformation then the spacetime must be flat.  The theorem is easily understood. For simplicity, consider the case of space not the case of spacetime. For better imagination, consider two dimensional space (surface) not three dimensional space. The simplest surfaces are the flat plane and the curved sphere surface. The quadratic form for plane $\xi \eta $ is $ds^2 = d\xi ^2 +d\eta  ^2$, which is exactly the Pythagoras theorem of right triangle. The quadratic form for sphere surface has a similar but definitely positive form to (5). However, it can never be transformed into the Pythagoras formula by 
whatever coordinate transformation.  In the case of homogeneous gravity, such coordinate transformation does exist which is the formula (7). I have another coordinate transformation in part (v) which cancels the homogeneous gravity too. Therefore, the spacetime $txyz$ which presents homogeneous gravity must be flat. Because $g_{\alpha \beta }$ in (9) is not the Minkowski metric $\eta _{\alpha \beta }$, the quantity $\bar s$ in (9) is not real distance along the curves of the flat spacetime $txyz$. Therefore, Einstein`s geometrization fails to the description of homogeneous gravity.
Because the quadratic forms (9) and (8) describe homogeneous gravity successfully, they initiate a method on gravitational study. The method abandons the geometrization of gravity and requires that spacetime be flat with its only geometric quantity being the Minkowski metric $\eta _{\alpha \beta }$. The quantity $\bar s$ is called effective distance and $g _{\alpha \beta }$ is called effective metric.  Both have no geometric meaning.        

{\bf (v) Square-root rescale which cancels homogeneous gravitational field.}
The only observable quantity in the static homogeneous gravitational field is the quadratic motion, $x=(1/2)gt^2 +x_1$. If we can find other coordinate transformation and the application of the above procedure leads to the same ``Pythagoras formula999 in the curvilinear coordinate system $\tau \xi $ and the same quadratic motion $x=(1/2)gt^2 +x_1$ in the rectangular coordinate system $tx$ then we can say that the new coordinate transformation cancels homogeneous gravity too. We try the following coordinate transformation,   
\begin{equation} 
\begin{array}{l}
\sigma =\sigma _0\sqrt{x/ x _0} , \\
\tau =t  
\end{array}
\end{equation}
where I introduce two constants $\sigma _0, x _0$ to fulfill the requirement that both $x$ and $\sigma $ have the same length unit. In the coordinate transformations provided in the following sections, if we do not see such constants then they are understood to have values of 1 and are not presented in the formulas for simplicity.  However, coordinate transformations are always understood to have homogeneous forms which are similar to the following 
\begin{equation} 
x =x _0 f(\xi /\xi _0).
\end{equation}
Substitution of the coordinate transformation (13) into the following ``Pythagoras formula999  
\begin{equation} 
d\bar s^2 =d\tilde \tau ^2-d\sigma ^2, 
\end{equation}
we have the following effective metric in $tx$ coordinate system,
\begin{equation} 
\begin{array}{ll}
d\bar s^2 &=d\tilde t^2-\frac{\sigma _0^2}{4x _0 x} dx^2 \\
     &=-g _{\alpha \beta }dx ^\alpha dx ^\beta
\end{array}
\end{equation}
where
\begin{equation} 
g_{00}=-1, g_{01}=g_{10}=0, g_{11}= \frac{\sigma _0^2}{4x _0 x}.
\end{equation}
The solution of the resulting geodesic equation (3) is any quadratic motion $x=(1/2)ht^2 +x_1 $ where $h$ is an arbitrary constant. That is, the coordinate transformation (13) cancels homogeneous gravity and particles experience no gravity in the curvilinear coordinate system $\tau \sigma $. However, $\tau \sigma $ is just a curvilinear coordinate space. It is not a reference frame  because the relation between $\sigma $ and $x$ is not linear. However, the coordinate system $\tau \xi $ (see (7)) is a global freely-falling frame by which people can make measurement. 

The coordinate system $\tau \sigma $ is a special curvilinear coordinate system. 
The main feature of the coordinate transformation (13) is that the space coordinate $\sigma $ is transformed to space coordinate $x$ independent of the  time coordinate transformation. From now on, our coordinate-rescales deal with spatial coordinates only. We consider $\sigma $ to be the curvilinear coordinate relative to the rectangular space $x$ and the transformation $\sigma =\sigma _0\sqrt{x/ x _0} $ is called an uneven rescale on the coordinate $x$. The coordinate space $ \sigma $ is called a shadow of the real space $x$ and $\sigma =\sigma _0\sqrt{x/ x _0} $ is called the shadow coordinate transformation. For any real point $x$, the radial line section from the origin to the point $\sigma _0\sqrt{x/ x _0}$ is called the shadow of the real section which is from the origin to the real point $x$. The former point is called the shadow point of the latter. In the following sections, the same definitions hold except that the coordinate $x$ is replaced by the radial coordinate $r$. However, I will not repeat the definitions. The distance between a point and its shadow can be large in the case of homogeneous gravitational field. This is understandable because there must exist infinite areas of mass distribution to maintain a mathematically homogeneous gravitational field. In section 4 we will see that the distance is small ($\simeq$ 3.0 km) for the gravitational field generated by solar mass.  

As argued in part (iv), if there is a global coordinate transformation which cancels gravity (i.\,e., the form (5) is transformed into (1)) then the spacetime must be flat (zero curvature). Its rectangular coordinate system must be $txyz$, and $\tau \xi \eta \zeta $ must be a curvilinear coordinate system of the spacetime. Both $\bar s$ and $g _{\alpha \beta }$ have no geometric meaning
because otherwise the rectangular coordinate system would be $\tau \xi \eta \zeta $ (flat spacetime with Minkowski metric) and a curvilinear coordinate system would be $txyz$. {\it Therefore, the gravitational theory which shares the properties of homogeneous gravity must be non-geometric.} Initiation of such theory is the goal of the present paper and a new equivalence principle is proposed as follows.

{\bf (vi) New equivalence principle (NEP).}
Spacetime is always flat with Minkowski metric $\eta _{ij}$. A tensor $g _{\alpha \beta } $ with Lorentz covariant symmetry is defined on the flat spacetime, which describes gravity and has no geometric meaning
\begin{equation}  
d\bar s^2 =-g _{\alpha \beta }dx ^\alpha dx ^\beta .
\end{equation}
The tensor $g _{\alpha \beta } $ is called effective metric. 
The effective distance $\bar s$ is not real distance on the spacetime $txyz$. 
Test particles follow the solution of the corresponding effective geodesic equation (3).  
The test particle`s motion extremizes the following functional variation 
\begin{equation}  
\delta \bar s=\delta \int ^{p_B}_{p_A}\frac {d\bar s}{dp}dp .
\end{equation}
For a galaxy, the gravitational redshift due to its mass distribution is not significant to be observed, i.\,e., $t \simeq \tau $. Furthermore, galactic gravitational fields are shown to be cancelled by spatial coordinate rescales. 
Therefore, a global spacetime coordinate transformation (12) 
can be found which transforms the quadratic form (18) into the following
\begin{equation} 
d\bar s^2 =d\tilde \tau ^2 -d\xi ^2 -d\eta ^2-d\zeta ^2.
\end{equation}
That is, the corresponding effective curvature is zero. This kind of effective metric (18) is called holonomic because the relation between $dx^{\alpha }$ and $d\xi ^{\alpha }$ (i.\,e., the equality of (18) to (20)), can be integrated into a global spacetime coordinate transformation (12). However, the gravitational field of an isolated point mass (e.\,g., a star, which is the basic component of galaxies), is non-holonomic because solar gravitational redshift is observed to be significant. That is, the corresponding effective curvature is non-zero.
The present paper shows that a nonholonomic spacetime boost transformation cancels the gravity.

Now we understand that the method of NEP is similar to the one of GR except that the spacetime of the former is flat while the one of the latter is curved. In NEP, therefore, the coordinate system $\tau \xi \nu \zeta $ is curvilinear and the inertial frame $txyz$ is the real rectangular coordinate system of the flat spacetime. The effective metric $g_{\alpha \beta }$ in NEP is a tensor field on the flat spacetime and measures the gravitational 999medium999 which is generated by the corresponding mass distribution. The 999medium999 curves the motion of (test) particles (i.\,e., extremizing effective distance $\bar s$) in the similar way the dielectric medium curves the propagation of light waves (extremizing refractive index $n$).
GR which attributes gravity to spacetime curvature, is actually based on the assumption that locally at each spacetime point there is a tangent flat Minkowski spacetime instead of a freely-falling one. 

{\bf (vii) Square-root, logarithmic, reciprocal, and translational rescales.}
We have already shown that homogeneous gravity is governed by the above NEP principle and the rescale is square-root. He (2005a) found that the gravitational field of 2-dimensional mass distributions of spiral galaxy disks can be derived by the principle too and the spatial radial coordinate rescale is logarithmic. A simple explanation of galactic rotation curves is given by the corresponding new stellar dynamics. The gravitational field of elliptical galaxies is possibly described by radial reciprocal rescale. In section 4, I will show that the gravitational field of an isolated point mass is also governed by the principle and is canceled by nonholonomic spacetime boost transformation. The radial translational rescale may contribute to the  cancellation too. 
The test particles follow the geodesic motion determined by the effective metrics and we have corresponding gravitational dynamics to all the cases. 
The corresponding effective curvature-tensor may be zero, i.\,e., holonomic (in the cases of homogeneous gravity and the gravitational fields of galaxies) or may be non-zero, i.\,e., nonholonomic (in the case of the gravitational field of an isolated point mass). However, the spacetime is always flat and $g_{\alpha \beta }$ has no connection to it.
 
{\bf (viii) Weakness of GR. }
Any inhomogeneous gravitational field can be considered to be static and homogeneous within small zone of spacetime. On the other hand, the geometrization of gravity (GR)  is claimed to be based on the cancellation of static homogeneous gravity by freely-falling frames. However, I have shown in part (iv) that geometrization fails to the description of homogeneous gravity. This is the main weakness of GR. 
Further more, the metric of the geometrization has to be determined by spacetime curvature and the corresponding Einstein equation is highly nonlinear and complicated. The common spatial parameters like distances and angles can not be computed directly from any coordinate system. They are determined by the metric because spacetime is curved.  

GR encounters many difficulties. Theoretically, the total gravitational energy  is not well defined and the gravitational field can not be quantized because it is connected to the space-time background itself. Realistically, Einstein equation permits very few metric solutions. Anisotropic and non-vacuum metric solutions which deal with 2-dimentional mass distributions like spiral galaxy disks do not exist in literature, to my knowledge. Astronomic observations reveal many problems which can not be resolved by GR and people resort to dark matter. 
Zhytnikov and Nester (1994)`s study indicates that   
the possibility for any geometrized gravity theory to explain the behavior of galaxies without dark matter is rather improbable. 
Therefore, looking for a non-geometrized yet relativistic gravitational theory of galaxies is of great interest. 
The present paper, following He (2005a), provides a preliminary theory of the kind. In the theory, static gravity can be cancelled by spatial coordinate rescales (holonomic transformation) or by nonholonomic boost.
Four types of rescales are found. The corresponding dynamical equations are ready for tests. 

GR is widely accepted because some of its calculation are testified by solar measurements. However, {\it the curvature of space-time was never measured and it is never proved that there exists no other dynamical equation similar to (3) whose solution gives the same or similar first-order predictions for solar system as GR.} The present paper shows the existence.   

Section 2 discusses general diagonal effective metric and the solution of its effective geodesic equation, taking spatial radial coordinate rescale as example.
Section 3 discusses the spatial logarithmic and reciprocal coordinate rescales   which cancel the gravitational fields of spiral galaxy disks and elliptical galaxies respectively.
Section 4 discusses nonholonomic boost transformation which cancels the gravitational field of isolated point mass.
The metric is different from the Schwarzschild one. To first order, its prediction on the deflection of light and the precession of the perihelia of the planetary orbits is the same as the one of GR.
Section 5 is conclusion.

\section{ General Discussion of Diagonal Effective Metric }

He (2005a) indicates that the logarithmic arms of ordinary spiral galaxies are the evidence that the gravitational field generated by the mass distribution of spiral galaxy disks can be canceled by the $t \xi \phi \theta $ coordinate system,
\begin{equation} 
\begin{array}{l}
t=t,               \\
\xi = \xi _0 \ln (r /r _0), \\
\phi =\phi ,     \\
\theta = \theta
\end{array}
\end{equation}
where $ r \phi \theta \,(0\leq \phi <2\pi , 0\leq \theta <\pi )
$ is the spherical polar coordinate system in the real rectangular $xyz$ space and $\xi _0 $, $r _0$ are constants. The $\xi \phi \theta $ coordinate system is simply the uneven rescale on the spatial radial lines in the $xyz$ space.
Because $\xi =p(r ) = \xi _0\ln (r /r _0)$, the rescale is called a logarithmic one.
The rescale and all others discussed in the following are the ones on the spatial radial lines in $xyz$ space. Therefore, we give the general result on spatial radial rescale $\xi  =p (r)$ in the present section.

{\bf (i) Effective metric and geodesic equation. } 
Let 
\begin{equation} 
\xi =p(r )
\end{equation} 
be spatial radial rescale.
Because the rescale cancels gravity, we have 
\begin{equation} 
\begin{array}{ll}
d\bar l^2&=d\xi ^2 +\xi ^2 (d\theta  ^2 +\sin ^2\theta d\phi ^2  ) \\
&=p^{\prime 2}(r ) dr ^2 + p^{ 2}(r ) ( d\theta  ^2 +\sin ^2\theta d\phi ^2  )
\end{array}
\end{equation} 
where $\bar l$ is the spatial effective distance in the rectangular space $xyz$. If the curvilinear coordinate $\xi \eta \zeta $ is imagined to be the rectangular one of an independent space then $\bar l$ is its real spatial distance and $(\xi , \theta ,\phi )$ its polar coordinates.  

The diagonal effective metric of the flat spacetime $txyz$ which describes the static radial gravitational fields must be the following
\begin{equation} 
\begin{array}{ll}
d\bar s ^2 &=B(r )d\tilde t^2-(p^{\prime 2}(r ) dr ^2 + p^{ 2}(r ) ( d\theta  ^2 +\sin ^2\theta d\phi ^2  ))\\
      &=B(r) d\tilde t^2-(A(r) dr ^2 + C( r) ( d\theta  ^2 +\sin ^2\theta d\phi ^2  ))\\
      &=-g_{\alpha \beta }dx^\alpha dx^\beta
\end{array}
\end{equation} 
where $dx^\alpha =(ct, r, \phi , \theta ) $ and  
\begin{equation} 
\begin{array}{l}
g_{00}(\equiv g_{tt})=-B( r),\\ 
g_{11}(\equiv g_{rr})= p^{\prime 2}(r ) =A( r), \\ 
g_{22}(\equiv g_{\theta  \theta  })= p^{2}(r ) =C( r), \\
g_{33}(\equiv g_{\phi  \phi  })= p^{2}(r )\sin ^2\theta , \\
g_{\alpha \beta } \equiv 0, \: \alpha \not= \beta.  
\end{array}
\end{equation} 
The coefficient $B(r )$ describes the gravitational redshift as suggested by GR.
All other coefficients are determined by the spatial radial rescale
 $\xi  =p (r)$. If $B(r )\equiv 1$ then the above quadratic form of effective metric can be transformed into the following 999Pythagoras formula999,
\begin{equation} 
d\bar s ^2 =d\tilde \tau^2-(d\xi ^2 +\xi ^2(d\theta ^2 +\sin ^2\theta d\phi ^2))
\end{equation} 
by a global spacetime coordinate transformation 
\begin{equation} 
t=\tau, r=r _0f(\xi /\xi _0 ), \phi =\phi ,\theta =\theta 
\end{equation} 
where $ r= r _0f(\xi /\xi _0 )$ is the inverse of the radial coordinate rescale $\xi =p(r )$.
This is true for homogeneous gravitational field and approximately true for the gravitational fields generated by the mass distributions of spiral galaxy disks and elliptical galaxies. However, it is not true for the gravitational field generated by a point mass $M$ where $B(r )=1-2MG/(c^2r)$. 

In spiral galaxy disks, stars are approximately planar motion. This is also true for any individual star of elliptical galaxies and any individual test particle in the gravitational field of an isolated point mass because the gravitational fields are described by the above effective metric.
Therefore, we take $\theta =$ constant $= \pi/2$ and our formulas involve three variables $t,r,\phi$ only,
\begin{equation} 
\begin{array}{ll}
d\bar s ^2 &= B(r) d\tilde t^2-(p^{\prime 2}(r ) dr ^2 + p^{ 2}(r )d\phi  ^2 ) \\
      &=B(r) d\tilde t^2-(A(r) dr ^2 + C( r) d\phi ^2  )\\
      &=-g_{\alpha \beta }dx^\alpha dx^\beta 
\end{array}
\end{equation} 
where $dx^\alpha =(ct, r, \phi )$ and  
\begin{equation} 
\begin{array}{l}
g_{00}(\equiv g_{tt})=-B( r),\\ 
g_{11}(\equiv g_{rr})= p^{\prime 2}(r ) =A( r), \\ 
g_{22}(\equiv g_{\phi  \phi  })= p^{2}(r )=C(r ) , \\
g_{\alpha \beta } \equiv 0, \: \alpha \not= \beta.  
\end{array}
\end{equation} 
Test particles move on curved orbits due to the effective metric $g_{\alpha \beta }$. 
Their motion follows the geodesic equation (3).
The only non-vanishing components of its affine connection are
\begin{equation} 
\begin{array}{ll}
\Gamma ^r_{rr} = \frac {A^\prime (r)}{2A(r)}, &
\Gamma ^r_{\phi  \phi  }=-\frac {C^\prime (r)}{2A(r)},\\
\Gamma ^r_{\tilde t \tilde t }= \frac {B^\prime (r)}{2A(r)},&
\Gamma ^\phi _{r \phi  }=\Gamma ^\phi _{ \phi  r}
=\frac {C^\prime (r)}{2C(r)},\\
\Gamma ^{\tilde t}_{r \tilde t }=\Gamma ^{\tilde t}_{ \tilde t r}=\frac {B^\prime (r)}{2B(r)}& 
\end{array}
\end{equation} 
where $ A^\prime (r) =dA (r)/dr$, etc.. 

{\bf (ii) Constants of the motion.}
{\it Note that the above formula (30) and the solution of its corresponding effective geodesic equation in the following (i.\,e., the formulas (32), (33), and (35)) hold to arbitrary diagonal effective metric (24) or (28) which is not necessarily a coordinate rescale. }  
The geodesic equation is solved by looking for constants of the motion.
In fact, the following solutions (32), (33) and (35) are standardized ones which can be found in, e.\,g., Weinberg (1972). The only difference is about $A(r ), B(r ), C(r )$. For example, $A(r )=1/B(r ), B(r )=1-2MG/(c^2r), C(r )=r^2$ is the Schwarzschild solution of Einstein`s geometrodynamics.   
I repeat Weinberg (1972)`s argument in deriving the solutions.  

The geodesic equations which involve $d^2\tilde t /dp^2$ and $d^2\phi /dp^2$ are called time component equation and polar-angle component equation respectively. They can be rewritten as the following, 
\begin{equation} 
\begin{array}{l}
\frac {d }{dp}(\ln \frac {dt }{dp}+\ln B( r) )=0, \\
\frac {d }{dp}(\ln \frac {d\phi }{dp}+\ln C( r) )=0. 
\end{array}
\end{equation} 
These yield two constants of the motion. The first one is absorbed into the definition of $p$. I choose to normalize $p$ so that the solution of the time component equation is
\begin{equation} 
\frac {dt }{dp}=1/ B( r). 
\end{equation}
The other constant is obtained from the polar-angle component equation,
\begin{equation} 
C( r)\frac {d\phi }{dp}=J. 
\end{equation}
The formula is used to study spiral galaxy rotation curves in the next section.
If $C( r)$ is $r^2$ as suggested by the Schwarzschild solution then $J$ is the conservative angular momentum per unit mass and the rotation speed is $rd\phi /dt=JB( r)/r$.  The Schwarzschild solution further suggests $B( r) \approx 1$ at large distance $r$ from the galaxy center and we expect a decreasing rotation curve, $V( r)= rd\phi /dt = J/r$. Real rotation curves are often constant over a large
range of radius and rise outwards in some way. In our proposition (the formula (21)), however, $C( r)=\xi _0^2\ln ^2(r/r _0), B( r)=1$ and we have a non-decreasing rotation curve.

Note that the common angular momentum is no longer conserved ($C(r ) \not= r^2$) in NEP theory, because of the radial coordinate-rescale cancellation of gravity. Instead, the shadow angular momentum is conserved whose direct result is that galactic rotation curves no longer decrease outward and no dark matter is required for their explanation (He, 2005a).

Furthermore, we have a gravitational dynamic equation which is the third component of the geodesic equation (the polar-distance component equation),
\begin{equation} 
\frac {d^2r }{dp^2}+ \frac {A^\prime }{2A}\left (\frac {dr }{dp}\right )^2-
\frac {C^\prime }{2A}\left (\frac {d\phi }{dp}\right )^2+
\frac {c^2B^\prime }{2A}\left (\frac {dt }{dp}\right )^2=0.
\end{equation} 
With the help of the other solutions, we have the last constant of the motion, 
\begin{equation} 
A( r)\left (\frac {dr }{dp}\right )^2+\frac {J^2 }{C( r)}-\frac {c^2}{B( r)}=-E\:({\rm constant}).
\end{equation} 

\section{New Stellar Dynamics of Galaxies }

{\bf (i) New Stellar dynamics of spiral galaxy disks.}
For the logarithmic rescale $\xi=\xi _0 \ln (r /r _0)$ (see (21)), we have
\begin{equation}
\begin{array}{l}
p(r )=\xi _0 \ln (r/r _0),                       \\
A(r )= p^{\prime 2}(r )= \xi _0^2/r^2 ,           \\
C(r )= p^{2 }(r )= \xi _0^2\ln ^2(r/r _0).
\end{array}
\end{equation} 
Because astronomic observation does not show any significant gravitational redshift due to galaxy mass distributions, it is good approximation to choose $B(r )=1$. Therefore, the effective metric for spiral galaxy disks is
\begin{equation}
d\bar s ^2 =d\tilde t^2-\left (\frac {\xi _0^2}{r^2} dr ^2 + \xi _0^2\ln ^2
\left (\frac {r}{r _0}\right ) d\phi ^2 \right ).
\end{equation}
The stellar dynamics is the solutions (32), (33), and (35) with the corresponding $A(r ), B(r ), C(r )$ being substituted. It is ready for test on galaxy observations.

{\bf (ii) Galaxy patterns (the origin of the coordinate rescale). } 
The coordinate rescale origins from the study of galaxy patterns (i.\,e., the light distributions $\rho (x, y)$). 
I proposed to use curvilinear coordinate systems to study galaxy light distribution patterns with the help of a symmetry principle (He, 2003). The light pattern of spiral galaxy disks is associated with an orthogonal curvilinear coordinate system $(\lambda , \mu )$ on the disk plane and the symmetry principle is that the components of the gradient vector $\nabla f(x, y)$ associated with the local reference system of the curvilinear coordinate lines depend on single curvilinear coordinate variables $\lambda $ and $\mu $ respectively, where $f(x,y)= \ln \rho (x,y)$.
The curvilinear coordinate system turns out to be the symmetrized one of the spatial part of the coordinate system (21). The coordinate system together with the symmetry principle determines the light distributions of spiral galaxy disks uniquely.
This method determines all regular galaxy patterns (He, 2005a and b).
The light distributions of arms can not be obtained in this manner. Arms are  density waves in the coordinate space (21) and destroy the above symmetry principle.  

{\bf (iii) Curved waves.}  
People generally consider harmonic plane waves respective to the real Cartesian coordinate system $(x,y)$ itself, $\cos (ax+by+ct)$, where $a,b,c$ are constants. The lines of wave crests (i.\,e., the lines parallel to $ax+by =$ constant at fixed time $t$)  cross any line of propagating direction at uniformly distributed points in the real space. Some people consider the harmonic waves respective to the polar coordinate systems $(r,\phi  )$, $\cos (ar+b\phi  +ct)$ whose lines of crests  are curved on the real spiral galaxy disk plane and cross any line of propagating direction on the plane at uniformly distributed points too. In fact, the lines of crests are $r \propto \phi  $ which express the unreal linear arms in spiral galaxies.
We know that ``free999 light waves (i.\,e., light propagation in vacuum) are straight while inhomogeneous dielectric medium curves the waves.  
Similarly, the density waves (arms) in spiral galaxies experience inhomogeneous gravitational fields and they have logarithmic curvatures: $\ln r \propto \phi  $, that is, the crest lines of the density waves cross any line of propagating direction at unevenly distributed points in the real space. Therefore, we need to rescale the radial lines from the galaxy centers, $\xi = \xi _0 \ln (r/r _0)$, to obtain a new coordinate systems $(u=\xi \cos \phi  ,v=\xi \sin \phi )$.
The harmonic plane waves 
\begin{equation} 
\cos (a\xi +b\phi  +ct),
\end{equation} 
respective to the new polar coordinate system $(\xi , \phi  )$ present logarithmic curvatures in real spaces of spiral galaxy disks, $\xi \propto \ln r \propto \phi  $. The crest lines (the arms) on the real spiral disk plane cross any line of traveling direction at unevenly distributed points.
Therefore, the waves which experience no gravity in the ``free-fall999 curvilinear coordinate system $\xi \phi $ are the physical waves which experience gravity in the real rectangular Cartesian space.

{\bf (iv) A model of galactic rotation curves. }
The solution (33) of the polar-angle component equation suggests a model of galactic rotation curves:  
\begin{equation} 
V(r)=r\frac {d\phi }{dt}=rJB(r)/C(r)
\end{equation}
with $B(r)\equiv 1$.  I review the model presented in He (2005a) in the following.
Except the constants of the motion, all other parameters from the effective metric of a specific galaxy disk are determined by the gravitational field of the background disk and are identical for all stars from the disk. The constant of motion $J$ has different values for different stars. But its averaged value $\bar J$ by all stars is a constant of the galaxy and does not depend on the radial distance $r$ of the galaxy. Finally we have a rotation curve model for spiral galaxy disks:  
\begin{equation}  
V(r)=r\frac {d\theta  }{dt}=r\bar J/(\xi _0^2\ln ^2(r/r _0)). 
\end{equation}
This rotation curve of pure spiral galaxy disks never decreases outwards. Instead, the model predicts a final rise of the curves at large distances from the galaxy centers. This is consistent to the astronomic observations of some galaxies. For the other galaxies, we need the rotation speed observations of large distances from the galaxy centers and compare the data with the prediction.  

For the model (40), we see a singularity near the galaxy centers, $V(r_0)=+\infty $. 
The calculation of Newtonian theory for pure galactic disks suggests a peaked but smooth rotation speed (Binney and Tremaine 1987; Courteau 1997).
My theory suggests a singularly peaked rotation curves for pure disk galaxies.
This indicates that my theory is a correction to Newtonian theory on spiral galaxies if we assume that there is no significant dark matter.
Real spiral galaxies, however, are always accompanied by 3-dimensional bulges near their centers.
We expect that the mass distributions of bulges pare off the singular peaks.
Therefore, we multiply a function $b(r)$ to the formula (40) to give account for the contribution of spiral galaxy bulges. Because the bulges have zero contribution at far distances from the galaxy centers, we require $b(r)\rightarrow 1$ for $r \rightarrow \infty $.
One of the simplest choices is the following  
\begin{equation} 
b(r)=(r-r_0)^2/(r\sqrt{(r-r_0)^2+c_0^2})
\end{equation}
where the numerator helps remove the singularity and the factor $r$ in the denominator results in a steep rise of the rotation curve near the galaxy center (a suggestion from the shapes of real rotation curves). 
The parameter $c_0$ determines the degree in which the singular peak is pared off. 
A subscript $b$ in the formula of the phenomenological model is used to indicate the bulge contribution besides the disk one,
 \begin{equation} 
\begin{array}{l}
V_b(r)=b(r)V(r)\\
=\bar J(r-r_0)^2/(\xi _0^2\ln ^2(r/r _0)\sqrt{(r-r_0)^2+c_0^2}). 
\end{array}
\end{equation}
The curve fits real rotation data (He, 2005a).

{\bf (v) Stellar dynamics of elliptical galaxies.}
As shown in the following, the radial gravitational field in 3-dimensional  elliptical galaxies has the reciprocal rescale $\xi=\xi_ 0r _0/r $. Therefore,
\begin{equation}
\begin{array}{l}
p(r )= \xi _0 r _0/r,                       \\
A(r )= p^{\prime 2}(r )= \xi _0^2 r _0^2/r^4,           \\
C(r )= p^{2 }(r )= \xi _0^2 r _0^2/r^2 .
\end{array}
\end{equation} 
Because astronomic observation does not show any significant gravitational redshift due to galaxy mass distributions, it is good approximation to choose $B(r )=1$. Therefore, the effective metric for elliptical galaxies is
\begin{equation}
d\bar s ^2 =d\tilde t^2-\left (\frac {\xi _0^2r^2 _0}{r^4} dr ^2 + \frac {\xi _0^2r _0^2} {r^2} d\phi ^2 \right ).
\end{equation}
The stellar dynamics is the solutions (32), (33), and (35) with the corresponding $A(r ), B(r ), C(r )$ being substituted. 

{\bf (vi) The origin of the reciprocal rescale.}
He (2005b) studied elliptical galaxy patterns by employing a curvilinear coordinate system $(\lambda , \mu ,\nu )$. The corresponding 3-dimensional spatial coordinate transformation is
\begin{equation} 
\left \{
\begin{array}{l}
\lambda = x /(x ^2 +y^2+z ^2) ,   \\
\mu = y /(x ^2 +y^2+z ^2) ,   \\
\nu = z /(x ^2 +y^2+z ^2) , \\
-\infty <x, y, z <+\infty . 
\end{array}
\right .
\end{equation}
With the help of the above-said symmetry principle and a cut-off method, 3-dimensional light patterns are achieved and their projected light distributions on the sky plane fit real elliptical galaxy images very well.
Therefore, it is good assumption that the above coordinate transformation cancels the gravitational field of elliptical galaxy mass distributions. That is, the effective metric of the coordinate space $(\lambda ,\mu ,\nu )$ is Pythagoras theorem:
\begin{equation}
\begin{array}{l}
d\lambda ^2 +d\mu ^2 +d\nu ^2 \\
= \frac {1}{r^4}dr^2 + \frac {1}{r^2}( d\theta  ^2 +\sin ^2\theta d\phi ^2  )
\end{array}
\end{equation}
where $(r, \theta ,\phi )$ is the common spherical polar coordinates in the rectangular coordinate space $(x,y,z)$.
The formula is the spatial part of (44) with the constants ignored,  which indicates a radial reciprocal rescale. The corresponding stellar dynamics is ready for test.

Because any 2-dimensional spiral galaxy disk is always accompanied by 3-dimensional elliptical bulge, the disk dynamics (37) must be combined with the elliptical dynamics (46) to study stellar motion in spiral galaxies.

{\bf (v) Gravitational dynamics and pattern dynamics.}
By now we have studied the ``gravitational999 dynamics of galaxies. Note that I added quotation marks to the word ``gravitational999 because I deduced the galactic dynamics without any consultation to Newtonian theory of gravity. What I employed is the result about galaxy patterns and the principle of interactions of finite speed $(\leq c)$, a principle of special relativity. 
In solar gravitational dynamics, we are never worried about the speed because the sun is the dominated cause and light $c$ can be considered to be infinite.   
In galactic scale systems, however, the speed is infinitesimal when compared with the spatial scales. All applications of Newtonian dynamics to galactic systems encounter difficulties and resort to dark matters. Rejecting the assumption of dark matters means giving up Newtonian dynamics in galaxy study. Therefore, I would call the galactic dynamics developed in the present paper the pattern dynamics rather than gravitational dynamics.
This bears a resemblance to statistical mechanics.      
Galaxies are the large-scale ensemble of stars in the similar way that gas is the macroscopic ensemble of atoms which demonstrates totally different thermal properties from the ones of single isolated ``cold999 atom. 

\section{ Nonholonomic Boost Transformation and the
 Gravitational Field Generated by an Isolated Point Mass }

{\bf (i) Nonholonomic boost transformation and the effective metric of an isolated point mass $M$.}
The main difference of the gravitational field of an isolated point mass from the one of galaxies is that the gravitational redshift due to galaxy mass distribution is not significant while the solar gravitational redshift is significant to be observed.
Because the time coefficient $g_{00}$ of the diagonal effective metric describes the redshift and is the function of spatial radial position, we expect that the corresponding effective curvature is nonzero. That is, there is no global spacetime coordinate transformation (12) which
transforms the effective metric into Minkowski form (i.\,e., nonholonomic).

We know that Lorentz boost transformation
\begin{equation}
\left \{
\begin{array}{ll}
x&=\frac {\xi - \beta \tilde \tau }{\sqrt{1-\beta ^2}},           \\
\tilde t&=\frac {\tilde \tau  - \beta \xi}{\sqrt{1-\beta ^2}},   \\
\beta &=v/c         
\end{array}
\right .
\end{equation} 
leads to the equations between coordinate differentials
\begin{equation}
\left \{
\begin{array}{ll}
d\tilde t &=\frac {d \tilde \tau }{\sqrt{1-\beta ^2}},           \\
d  l&=d \bar l \sqrt{1-\beta ^2 }     
\end{array}
\right .
\end{equation} 
which are the time dilation and length contraction respectively. In the formulas, $\tilde \tau \bar l$ is the frame in which clock and length are at rest.  
This equation system of differentials, (48), is holonomic because it can be integrated and the resulting coordinate transformation is (47). The holonomic  boost (48) (or (47)) has a boost direction which is the $x$-axis.
Now I propose a boost which is nonholonomic and boosts to all spatial directions
\begin{equation}
\left \{
\begin{array}{ll}
d \tilde t &=\frac {d \tilde \tau }{\sqrt{1-2r_g/r}},           \\
d l&=d \bar l \sqrt{1-2r_g/r}     
\end{array}
\right .
\end{equation} 
where $r_g =GM/c^2$ and
\begin{equation}
d l^2= dx^2+dy^2+dz^2 = dr^2+ r^2(d\theta ^2+\sin ^2\theta d\phi ^2)      
\end{equation} 
is the spatial square distance of arbitrary direction (not necessarily the radial direction) in the real inertial spacetime $txyz$ while $d\bar l$ is the spatial distance in the curvilinear spacetime $\tau \xi \eta \zeta $ which does not present gravity. The formula (49), which resemble (48) with $2r_g/r$ playing the role of $\beta =v/c$, is considered to be a generalized boost transformation. However, it is different from the Lorentz boost. Firstly, Lorentz boost (48) has a boost direction ($x$-axis) and is holonomic (the resulting Lorentz transformation (47)) while the generalized one is nonholonomic. That is, (49) can not be integrated to give a global coordinate transformation (12). Secondly, Lorentz boost is the relationship between two inertial frames and is symmetric about the two frames while the generalized boost is the relationship between the inertial frame $txyz$ and the curvilinear coordinate system $\tau \xi \eta \zeta $. However, the generalized boost cancels the gravity of isolated point-mass.
The effective metric of the isolated point mass $M$ is 
\begin{equation} 
\begin{array}{ll}
d\bar s^2 &=d\tilde \tau ^2 -– 
d\bar l^2 = (1-2r_g/r) d \tilde t^2  –- \frac {dl^2}{1-2r_g/r}    \\
&= (1-2r_g/r) d \tilde t^2  

–- 
\left (\frac {dr^2}{1-2r_g/r}+\frac {r^2}{1-2r_g/r}( d\theta ^2+\sin ^2\theta d\phi ^2)\right )           \\
&=B(r )d \tilde t^2 -– 
(A(r )dr^2+ C(r )(d\theta ^2+\sin ^2\theta d\phi ^2))           
\end{array}
\end{equation}
where $(r, \theta ,\phi )$ is the common spherical polar coordinates in the rectangular coordinate space $(x,y,z)$ and 
\begin{equation}
\begin{array}{l}
B(r )= 1-2r_g/r,                       \\
A(r )= \frac{1}{1-2r_g/r } ,           \\
C(r )= \frac{r^2}{1-2r_g/r }  .
\end{array}
\end{equation} 
Note that $B(r )$ and $A(r )$ are exactly the ones from Schwarzschild metric in GR. However, $C(r )(=r^2/(1-2r_g/r)\simeq (r+r_g)^2 )$ is different, which means that the common angular momentum is not conserved. Instead, the shadow angular momentum is conserved. Because the distance between a point and its shadow is $2r_g= 2GM/c^2$ ($\approx $ 3.0km for the mass of sun (GR suggests a distance of $r_g$)), the phenomenon can be resolved only by high precision solar observation.

{\it Note that all solar GR tests were made on its first order (in $r_g/r$) predictions. Therefore, if $B(r ), A(r )$, and $C(r )$ in (51) all have the same   
first order (in $r_g/r$) approximations as GR then the corresponding effective metric gives the same predictions as GR, i.\,e., the same predictions of 
the deflection of light by the sun, the precession of perihelia, and the excess delay of radar echo near the sun. } In the remaining parts of the present section I verify the assertion and provide new results.

{\bf (ii) Effective metrics which generalize Newtonian theory and fulfill GR`s first-order prediction.}
We have shown that geometrization of gravity is not needed. While geometrization requires spacetime curvature to determine gravitational metric (Einstein equation), the effective metric of NEP has no such constraint. Without the constraint, we are free to see which kinds of effective metrics give exactly the same first order predictions as GR.
Here I present examples of such effective metrics.
Firstly, $B(r )$ in (52) and (51) can be any one of the following
\begin{equation}
B_1:\: 1-2r_g/r;\: B_2:\: (1-r_g/r)^2;\: B_3:\: \frac{1}{1+2r_g/r};\: B_4:\:\exp(-2r_g/r);\:\cdots 
\end{equation} 
where $r_g=GM/c^2,\;G$ is gravitational constant, $M$ the isolated point mass, and $c$ light speed.
Secondly, $A(r )$ can be any one of the following
\begin{equation}
A_1:\: \frac{1}{1-2r_g/r};\: A_2:\: (1+r_g/r)^2;\: A_3:\: 1+2r_g/r;\:  A_4:\: \exp(2r_g/r);\:\cdots . 
\end{equation} 
The $A(r )$ and $B(r )$ work because they have the same   
first order (in $r_g/r$) approximations as the Schwarschild ones in GR. I have further result. I will show that arbitrary function $C(r )$ together with the above $A(r )$ and $B(r )$ give exactly the same predictions of the deflection of light by the sun and the precession of perihelia. But the prediction on the excess radar echo delay depends on the choice of $C(r )$, i.\,e., depends on its first order approximation in $r_g/r$. 
Therefore, in the following examples of $C(r )$  
\begin{equation}
C_1:\: r^2;\: C_2:\: \frac{r^2}{1-2r_g/r};\: C_3:\: r^2(1+jr_g/r)^2;\: C_4:\:  r^2\exp(jr_g/r);\:\cdots  
\end{equation} 
where $j (\not= 0)$ is arbitrary constant, only the first one gives the same first order prediction on excess radar echo delay as GR. The nonholonomic boost metric corresponds to $j=1$ ($C_2$).

For verification of the assertion, following the formulas of Weinberg (1972) saves much time. 
Therefore, the following notes need to be read with the book and any three number set with parentheses in the following, e.\,g., (8.5.6), refers to the formulas in the book.
The sole difference of NEP calculation from GR is to change $r^2$ to $C(r )$ in the formulas. For simplicity, I choose $C(r )$ to be $C_3$ in (55)
\begin{equation}
C(r )=r^2(1+jr_g/r)^2
\end{equation} 
to verify the assertion. 
To begin, I note that 
the roughest approximation to the dynamical equations (32), (33), and (35) depends only on $B(r )$ because the only term which involves the large number $c$ is the third term in (35). Because all $B_i$ in (53) have the same first-order approximation 
\begin{equation}
1-2r_g/r,
\end{equation} 
this approximation of the dynamical equations gives
\begin{equation}
\begin{array}{c}
r^2 \frac{ d\phi }{dt} \simeq J                      \\
\frac{1}{2} (\frac{ dr }{dt} )^2 +\frac {J^2}{2r^2}-\frac{GM}{r} \simeq \frac{c^2-E}{2}
\end{array}
\end{equation} 
which are the formulas between (8.4.20) and (8.4.21) in Weinberg (1972).
These are the same equations of the Newtonian theory with $(c^2-E)/2$ playing the role of energy per unit mass. Therefore, $B(r )$ alone decides if the dynamics contains Newtonian theory as limiting case.  

{\bf (iii) First-order (in $r_g/r_0$) prediction in the deflection of light by the sun.}
Inserting the formulas (33), (53), (54), and (56) into (35) gives test particle`s orbital motion around an isolated point mass $M$ (the formula (8.4.30) in Weinberg (1972)),
\begin{equation}
\phi (r)=\pm\int \frac {A^{1/2}(r ) dr}{C(r ) \left ( \frac {1}{J^2B(r )}- 
\frac {E}{J^2}-\frac{1}{C(r )}\right )^{1/2}}. 
\end{equation} 
Formula (8.5.6) is the following with $E=0$ in the case of light,
\begin{equation}
\phi (r)-\phi (\infty )=\int ^\infty _rA^{1/2}(r )\cdot \left ( \frac {C(r )}{C(r_ 0)} 
\frac {B(r_0 )}{B(r)}-1\right )^{-1/2}\cdot \frac{1}{\sqrt{C(r )}}dr .
\end{equation} 
The argument of the second square root, in the first order approximation of $r_g/r_0$ where $r_0$ is the closest approach to the central isolated point-mass (the sun), is
\begin{equation}
\frac {C(r )}{C(r_ 0)}\frac {B(r_0 )}{B(r)}-1=
\left (\frac {r^2}{r_ 0^2}-1\right )\left (1- \frac {2 (1+j)r_g r}{r_0(r+r_0)}+\cdots \right ).
\end{equation} 
Therefore,
\begin{equation}
\begin{array}{ll}
\phi (r)-\phi (\infty )&=\int ^\infty _r \frac {dr}{r\sqrt{r^2/r_ 0^2-1} }
\left (1+\frac{(1-j)r_g }{r}+ \frac { (1+j)r_g r}{r_0(r+r_0)} +\cdots \right )\\
&=\sin^{-1}\left ( \frac {r_ 0}{r}\right )+\frac {r_ g}{r_0}\left ((1-j)+(1+j)-(1-j) \sqrt{1-\frac{r_0^2}{r^2}}-(1+j) \sqrt{\frac{r-r_ 0}{r+r_ 0}}\right )+\cdots
\end{array}
\end{equation}
We see a cancellation of $j$ on the first two terms after $r_g/r_0$ in the last result (comparing (8.5.7)). Therefore, the deflection of the orbit from a straight line
\begin{equation}
\Delta \phi =2|\phi (r_ 0)-\phi _\infty |-\pi =\frac{4r_ g}{r_ 0}
\end{equation}
remains unchanged. I have proved the first part of my assertion.

{\bf (iv) First-order (in $r_g/r_0$) prediction in the precession of perihelia.}
The formula (8.6.3) is 
\begin{equation}
\begin{array}{l}
\phi (r)-\phi (r_- )=\int ^r_{r_-}dr A^{1/2}(r )C^{-1}(r ) \\
\cdot \left ( \frac { C(r_-)(B^{-1}(r )-B^{-1}(r_- ))- C(r_+)(B^{-1}(r )-B^{-1}(r_+ )) 
}
 { C(r_+)C(r_-)(B^{-1}(r_+ )-B^{-1}(r_- )) } -\frac{1}{C(r )}\right )^{-1/2}.
\end{array}
\end{equation}
We consider the first-order approximation of the argument of the second square root. Note that
\begin{equation}
B^{-1}(r ) = \bar B^{-1}(u ) =1+\frac{2r_g}{u}+\frac{2(2+j)r^2_g}{u^2} +\cdots 
\end{equation}
where $u=r+jr_g $.
Because the first terms $1$ in (65) are completely canceled in (64), the second-order approximation of $\bar B^{-1}(u )$ is needed to achieve a first order approximation of the argument.
Note that $C(r )=u^2$. Therefore,
the first order approximation of the argument is an algebraic polynomial of $1/u$ of order two. Furthermore, it vanishes at $u_{\pm }(=r_{\pm }+jr_g )$, so  
\begin{equation}
\begin{array}{l}
 \frac { u_-^2(\bar B^{-1}(u )- \bar B^{-1}(u_- ))- u_+^2(\bar B^{-1}(u )- \tilde B^{-1}(u_+ ))}
 { u_+^2u_-^2(\bar B^{-1}(u_+ )- \bar B^{-1}(u_- )) } -\frac{1}{u^2} \\
= D\left ( \frac {1}{u_-}-\frac {1}{u} \right )\left (\frac {1}{u}- \frac {1}{u_+} \right )
\end{array} 
\end{equation}
The constant $D$ can be determined by letting $u \rightarrow \infty $:
\begin{equation}
D= \frac { u_+^2(1-\bar B^{-1}(u_+ ))- u_-^2(1-\bar B^{-1}(u_- ))}
 { u_+u_-(\bar B^{-1}(u_+ )- \bar B^{-1}(u_- )) }.
\end{equation}
Because $1/u=1/r -jr_g/r^2$ and $jr_g/r^2 \ll r_g/r_0$, we have, in first order approximation of $r_g/r$, 
\begin{equation}
D=1-(2+2j)r_g\left ( \frac{1}{r_+}+ \frac{1}{r_-}\right ).
\end{equation} 
Similarly, the factor on the right hand side of (66) can be approximated by

\begin{equation}
\left ( \frac {1}{u_-}-\frac {1}{u} \right )\left (\frac {1}{u}- \frac {1}{u_+} \right )\simeq
\left ( \frac {1}{r_-}-\frac {1}{r} \right )\left (\frac {1}{r}- \frac {1}{r_+} \right ).
\end{equation}
By introducing a new variable $\psi $, we have finally
\begin{equation}
\begin{array}{ll}
\phi(r )-\phi(r_-)&\simeq \left [1+\frac{1}{2}(2+2j)r_g \left ( \frac{1}{r_+}+ \frac{1}{r_-}   \right ) \right ]\int^r_{r_-}\frac {(1+(1-2j)r_g /r)dr }{
r^2 [( \frac {1}{r_-}-\frac {1}{r} )(\frac {1}{r}- \frac {1}{r_+} )]^{1/2}     }\\
&=[1+\frac{1}{2}(2+2j +1-2j)r_g ( \frac{1}{r_+}+ \frac{1}{r_-}   )][\psi +\frac{\pi }{2} ]\\
&-\frac{1}{2}(1-2j)r_g ( \frac{1}{r_+}- \frac{1}{r_-}   )\cos \psi .
\end{array} 
\end{equation}
Again we see a cancellation of $j$ in the final result and the precession in one revolution remains unchanged
\begin{equation}
\Delta \phi =\frac {6\pi GM}{L} \; {\rm radians/revolution }
\end{equation} 
where
\begin{equation}
\frac{1}{L}=\frac{1}{2}\left ( \frac{1}{r_+}+ \frac{1}{r_-} \right )  .
\end{equation} 

{\bf (v) First-order (in $r_g/r_0$) prediction in the radar echo excess delay.}
However, $j$ is not cancelled in the formula of excess radar echo delay and the first-order approximation in $r_g/r_0$ is different from that of GR. 
The formula of the time required for light to go from $r_0$ to $r$ or from $r$ to $r_0$ is (8.7.2)
\begin{equation}
t(r, r_0)=  \frac {1}{c}\int ^r_{r_0} \left (\frac{A(r )/B(r ) }{
1- \frac {B(r )}{B(r_0)} \frac {C(r_0 )}{C(r)} } \right )^{1/2} dr .  
\end{equation}
In first-order approximation, the denominator is
\begin{equation}
1- \frac {B(r )}{B(r_0)} \frac {C(r_0 )}{C(r)} 
\simeq \left (1- \frac {r_0^2}{r^2}\right )\left (1-\frac {2(1+j)r_g r_0}{r(r+r_0)}\right ).  
\end{equation}
Therefore,
\begin{equation}
\begin{array}{ll}
t(r, r_0)&\simeq  \frac {1}{c}\int ^r_{r_0}\left (1-\frac {r_0^2}{r^2}\right )^{-1/2}\left (1+\frac{2r_g}{r} + \frac{(1+j)r_gr_0}{r(r+r_0)} \right)dr\\
        &= \left (\sqrt{r^2-r^2_0}+ 2r_g\ln\frac{r+\sqrt{r^2-r^2_0} }{r_0}
+(1+j)r_g (\frac{r-r_0}{r+r_0})^{1/2}\right )/c.
\end{array}
\end{equation} 
The leading term $\sqrt{r^2-r_0^2}/c$ in the formulas is what we should expect if light traveled in straight lines at its speed $c$. The remaining terms in the formulas produce NEP delay and GR $(j=0)$ delay respectively in the time it takes a radar signal to travel to Mercury and back. These excess delays are maximums when Mercury is at superior conjunction and the radar signal just grazes the sun.
Detailed calculation (Weinberg, 1972) shows that the maximum round-trip excess  delay  is 
\begin{equation}
(\Delta t)_{{\rm max }}\simeq 19.7 (1+j+11.2) \; \mu \,{\rm sec }.
\end{equation} 
The nonholonomic boost metric corresponds to $j=1$ and its prediction is 
$(\Delta t)_{{\rm max }}\simeq 260 \mu \,{\rm sec } $ while the corresponding result of GR $(j=0)$ is 240 $\mu$\,sec. The difference is less than 8 percents.

However, there is difficulty in their tests.
We can transmit radar signals to Mercury at its series of orbital positions around the event of superior conjunction. The time for single round-trip is many minutes and an accuracy of the order of 0.1 $\mu $\,sec can be achieved (Anderson {\it et al.}, 1975). In order to compute an excess time delay, we have to know the time $t_0$ that the radar signal would have taken in the absence of the sun`s gravitation to that accuracy. This accuracy of time corresponds to an accuracy of 15 meters in distance. This presents the fundamental difficulty in the above test. In order to have a theoretical value of $t_0$, Shapiro`s group proposed to use GR itself to calculate the orbits of Mercury as well as the earth (Shapiro, 1964; Shapiro {\it et al.}, 1968; Shapiro {\it et al.}, 1971). The data of time for the above series of real round-trips minus the corresponding theoretical values of $t_0$ presents a pattern of excess time delay against observational date and was fitted to the excess delay calculated by the formula (75) with a fitting parameter $\gamma $.
The group and the following researchers found that GR, among other similar theories of gravity represented by $\gamma $, fits the pattern best.

Now I have a NEP gravitational theory of nonholonomic boost. Because angular momentum is not conserved, the orbits and orbital motion are much different.  It would be very interesting to use the new theory to test the same radar echo data. That is, we use the new gravitational dynamics to calculate the orbits of Mercury and the earth to achieve the above-mentioned theoretical values $t_0$. Similarly we add the same parameter $\gamma $ to the nonholonomic boost metric (52) and obtain a corresponding formula to (75). If the NEP metric fits the above-said pattern best then we can say that it is a competing theory to GR. However, the actual values of excess time delays are as difficult to resolve as the shadow angular momentum conservation of solar mass.      

{\bf (vi) Recovery of Schwarzschild metric.}
We can recover Schwarzschild metric by combining a spatial radial translational coordinate rescale to the above nonholonomic boost transformation (49).
That is, $txyz$ in (49) is considered to be curvilinear coordinate system. The real rectangular coordinate system in an inertial frame is $TXYZ$ and the spatial radial translational coordinate rescale is $r=p(R )=R+mr_g$ where $m$ is another arbitrary constant.
Therefore, $ B_2(R )=1,\, A_2(R )= p^{\prime 2}(R )= 1,\, C_2(R )= p^{2 }(R )= (R+mr _g)^2$.
Combined with the nonholonomic boost, $ B_1(r )=1-2r_g/r,\, A_1(r )=1/(1-2r_g/r),\, C_1(r )=r^2/(1-2r_g/r)$,
finally we have
\begin{equation}
\begin{array}{ll}
d\tilde \tau ^2&=B_1(r )d\tilde t^2 \simeq (1-2 r_g/R)d\tilde T^2,               \\
d\bar l^2 &=A_1(r )dl^2\simeq \frac {dl^2 }{1-2 r_g/R}           \\
&=\frac {dR^2 }{1-2 r_g/R}+\frac {(R+mr_g)^2 }{1-2 r_g/R}( d\theta ^2+\sin ^2\theta d\phi ^2) .            
\end{array}
\end{equation} 
Therefore,
\begin{equation}
\begin{array}{l}
B(R )\simeq 1-2 r_g/R,               \\
A(R )\simeq \frac {1 }{1-2 r_g/R},           \\
C(R )\simeq (R+2(m+1)r_g )^2.            
\end{array}
\end{equation} 
It can be seen that Schwarzschild metric is recovered when choosing $m=-1$.

\section{ Conclusion }

It is shown that Einsten`s geometrization fails to describe homogeneous gravity. A new equivalence principle (NEP) is proposed. 
Spacetime is always flat with Minkowski metric. A tensor $g _{\alpha \beta } $ which is covariant with respect to all curvilinear coordinate transformations (including non-curvilinear Lorentz transformations) is defined on the flat spacetime, which describes gravity and has no geometric meaning.
Test particles follow the solution of the corresponding effective geodesic equation. Static gravity can be canceled by holonomic or nonholonomic coordinate transformations. 

I proposed galaxy pattern dynamics which explains galaxy rotation curves successfully.
That is, I deduced the stellar dynamics without any consultation to Newtonian gravitational dynamics. Instead, I employed the results about galaxy patterns and the principle of interactions of finite speed $(\leq c)$, a principle of special relativity. 
In solar gravitational dynamics, we are never worried about the speed $c$ because it can be considered to be infinite.   
In galactic scale systems, however, the speed is infinitesimal compared with the large spatial scales.
All applications of Newtonian dynamics to galactic systems encounter difficulties and resort to dark matters. Rejecting dark matters means the abandonment of Newtonian dynamics in galaxy study. Therefore, I would call the galactic dynamics developed in the present paper the pattern dynamics rather than gravitational dynamics.
This bears a resemblance to statistical mechanics.      
Galaxies are the large-scale ensemble of stars in the similar way that gas is the macroscopic ensemble of atoms which demonstrates totally different thermal properties from the ones of single isolated ``cold999 atom. 

Also I proposed to use nonholonomic boost to generalize Newtonian dynamics. 
This generalization shares the properties of Lorentz transformation, which should help quantize gravity.
I explored its classical solar application.
The corresponding effective metric is different from that of Schwarzschild. To first order, its prediction on the deflection of light and the precession of the perihelia of planetary orbits is the same as the one of general relativity (GR). 
Its further implication is left for future work.

\end{document}